\documentclass{aa}  
\usepackage{graphicx}
\usepackage{longtable,lscape}
\usepackage[authoryear]{natbib}
\bibpunct{(}{)}{;}{a}{}{,}
\usepackage{txfonts}
\begin{document}
  \title{Foregrounds for observations of the cosmological 21~cm line: II. Westerbork observations of the fields around 3C196 and the North Celestial Pole} 
\author{G.~Bernardi\inst{1,2} \and A.G.~de~Bruyn\inst{1,3} \and G.~Harker\inst{1} \and M.A.~Brentjens\inst{3} \and B.~Ciardi\inst{4} \and V.~Jeli{\'c}\inst{1} \and L.~V.~E.~Koopmans\inst{1} \and P.~Labropoulos\inst{1} \and A.~Offringa\inst{1}\and V.N.~Pandey\inst{1} \and J.~Schaye\inst{5} \and R.M.~Thomas\inst{1,6} \and S.~Yatawatta\inst{1} \and S.~Zaroubi\inst{1}}
   \institute{Kapteyn Astronomical Institute, University of Groningen, PO Box 800, 9700 AV Groningen, The Netherlands\\
              \email{bernardi@astro.rug.nl}
         \and Harvard-Smithsonian Center for Astrophysics, 60 Garden Street, Cambridge, MA 02138, USA \and ASTRON, PO Box 2, 7990 AA Dwingeloo, The Netherlands \and Max-Planck Institute for Astrophysiscs, Karl-Schwarzschild-Stra\ss e, 1, 85748 Garching, Germany \and Leiden Observatory, Leiden University, PO Box 9513, 2300 RA Leiden, The Netherlands \and Institute for the Mathematics and Physics of the Universe (IPMU), The University of Tokyo, Chiba 277-8582, Japan} 
   \date{Received }

 
  \abstract{In the coming years a new insight into galaxy formation and the thermal history of the Universe is expected to come from the detection of the highly redshifted cosmological 21~cm line.} 
  {The cosmological 21~cm line signal is buried under Galactic and extragalactic foregrounds which are likely to be a few orders of magnitude brighter. Strategies and techniques for effective subtraction of these foreground sources require a detailed knowledge of their structure in both intensity and polarization on the relevant angular scales of 1-30 arcmin.} 
  {We present results from observations conducted with the Westerbork telescope in the 140-160~MHz range with 2~arcmin
  resolution in two fields located at intermediate Galactic latitude, centred around the bright quasar 3C196 and the
  North Celestial Pole. They were observed with the purpose of characterizing the foreground properties in sky areas
  where actual observations of the cosmological 21~cm line could be carried out. The polarization data were analysed through
  the rotation measure synthesis technique. We have computed total intensity and polarization angular power spectra.} 
  {Total intensity maps were carefully calibrated, reaching a high dynamic range, 150000:1 in the case of the 3C196 field. 
  No evidence of diffuse Galactic emission was found in the angular power spectrum analysis on scales smaller than
  $\sim$10~arcmin in either of the two fields. On these angular scales the signal is consistent with the classical
  confusion noise of $\sim$3~mJy~beam$^{-1}$. On scales greater than 30~arcmin we found an excess of power attributed to the Galactic foreground with an rms of 3.4~K and 5.5~K for the 3C196 and the NCP field respectively. The intermediate angular scales suffered from systematic errors which prevented any detection.

  Patchy polarized emission was found only in the 3C196 field whereas the polarization in the NCP area was essentially due to radio frequency interference. The polarized signal in the 3C196 field is close to the thermal noise for angular scales smaller than $\sim$10~arcmin. On scales greater than 30~arcmin it has an rms value of 0.68~K. The polarized signal appears mainly at rotation measure values smaller than 4~rad~m$^{-2}$.}
  {In regard of the detection of the cosmological 21~cm line, we conclude that Galactic total intensity emission
  lacks small-scale power, which is below the confusion noise level at the angular resolution of 2~arcmin. Galactic polarization, given its relative weakness and its small rotation measure values, is less severe than expected as a contaminant of the cosmological 21~cm line.} 


   \titlerunning{WSRT observations of Galactic foregrounds at 150~MHz}
   \maketitle
%

\section{Introduction}

Several facilities like GMRT\footnote{Giant Metrewave Telescope, http://www.gmrt.ncra.tifr.res.in}, LOFAR\footnote{Low Frequency Array, http://www.lofar.org}, MWA\footnote{Murchinson Widefield Array, http://haystack.mit.edu/ast/arrays/mwa}, 21CMA\footnote{21 Centemeter Array, http://web.phys.cmu.edu/$\sim$past} and PAPER\footnote{Precision Array to Probe the EoR, http://www.astro.berkeley.edu/$\sim$dbacker/eor} will aim to detect the redshifted 21~cm line from the epoch of reionization (EoR) in the near future. 

From observations of the cosmic microwave background polarization (Komatsu et al. 2009) and from observations of high redshift quasars (Becker, Rauch \& Sargent 2007) the Universe is expected to be reionized at $6 < z < 12$. 

Simulations of the evolution of the intergalactic medium also show that the EoR signal is expected to appear between 100 and 200~MHz and it is expected to have a strength of a few millikelvin with significant fluctuations with redshift (Ciardi \& Madau 2003; Furlanetto, Sokasian \& Hernquist 2004; Mellema et al. 2006; Thomas et al. 2008; Iliev et al. 2008). Therefore the presence of Galactic and extragalactic foregrounds, which are orders of magnitude higher than the cosmological signal (Shaver et al. 1999), is the most serious challenge for the measurement of the EoR signal. 

Several authors have studied the foreground contamination of the EoR signal together with the problem of removing it
through various techniques (Di Matteo, Ciardi \& Miniati 2004; Morales \& Hewitt 2004; Santos, Cooray \& Knox 2005, Morales, Bowman \& Hewitt 2006;
Wang et al. 2006, McQuinn et al. 2006; Gleser et al. 2008; Jeli{\'c} et al. 2008; Bowman, Morales \& Hewitt 2009;  Liu et al. 2009a; Harker et al. 2009b; Liu et al 2009b). Since none of these methods is fully blind but includes assumptions about the foreground properties, a further observational characterization of foregrounds is mandatory in order to improve these strategies.

Observational data aimed at characterizing the foreground properties are still lacking. All previous observations were conducted either at frequencies higher than 350~MHz or at low resolution. The characterization of the foreground properties from existing data was therefore quite uncertain (see Bernardi et al. 2009 for a discussion about foreground estimates at low frequencies from existing data).

Due to this uncertain knowledge, we have started an observational programme aimed at investigating the foreground properties for EoR observations at 150~MHz with the Low Frequency Front Ends (hereafter LFFE) on the Westerbork telescope (hereafter WSRT). Three different sky areas were targeted. In a previous paper we presented the observations of the Fan region, an area centred at Galactic latitude $b = 8^\circ$ (Bernardi et al. 2009, hereafter B09). Those data gave the first indication of the strength and the spatial properties of the foregrounds at frequency and angular scales relevant for the detection of the cosmological 21~cm line.

Those results could not, however, be extrapolated straightforwardly to sky areas at intermediate Galactic latitude where the actual EoR observations will be carried out. In particular, the Fan region showed bright polarized emission which could not be taken as representative for the polarization out of the Galactic plane.

In this paper we present the results from the observations of the other two fields, located at intermediate Galactic latitudes. The first field is centred on the very bright radio quasar 3C196, in one of the coldest regions of the Galactic halo. The second field is located close to the North Celestial Pole (NCP), which represents an ideal target to observe the cosmological 21~cm line because it would allow the collection of night time observations throughout the whole year at the geographic latitude of the LOFAR array (+53$^\circ$).

The paper is organized as follows: in Section~\ref{obs_res} we present the observations together with the data reduction and the initial results, in Section~\ref{pol_an} we analyse the polarized emission, in Section~\ref{power_spec} we present the power spectrum analysis and in Section~\ref{concl} we conclude.

\section{Observations and data reduction}
\label{obs_res}

The observational setup of the 3C196 and NCP fields was the same as the observations as the Fan region (B09). The observations of the two fields took place between the end of November and the beginning of December 2007 for a total of $6 \times 12$~hours for each field. The NCP field could be completely observed during night time while the last thirty to forty minutes of every session of the 3C196 field observations took place when the Sun had already risen.

The WSRT consists of 14 dishes of 25~m diameter each, ten of which (labelled  0 to 9) are on fixed locations 144~m apart. The other four (labelled  A to D) are movable on a rail track. The redundant baselines were not included in the imaging process because they cause very strong grating lobes. Combinations between the ten fixed antennas and the four movable ones were used to image the sky. Six array configurations were used, with the four movable dishes being incrementally moved by 12~m. This generated a uniform $uv$ coverage from 36~m up to the maximum spacing of 2760~m. Antenna 5 was missing in all the observations and this caused two gaps in the $uv$ plane at $|{\bf u}| \sim 300$ and $|{\bf u}| \sim 700$, where $|{\bf u}|=\sqrt{u^2+(v/\sin\delta)^2}$ is the projected distance in the $uv$ plane measured in wavelengths.
\begin{table}	
\caption[]{Summary of the observational setup.} 
\label{obs_table_3C196}
\begin{tabular}{l l}        
\hline\hline 
   
Coordinates of the 3C196 field 		& \\
centre (J2000.0) 			& $\alpha = 8^{\rm h}13^{\rm m}36^{\rm s}.0$, $\delta = 48^\circ 13' 03''$\\
Galactic coordinates 			& $l \simeq 171^\circ$, $b \simeq 33^\circ$\\
Coordinates of the NCP field 		& \\
centre (J2000.0) 			& $\alpha = 4^{\rm h}00^{\rm m}00^{\rm s}.0$, $\delta = 88^\circ 00' 00''$\\
Galactic coordinates 			& $l \simeq 124^\circ$, $b \simeq 25^\circ$\\
Number of spectral bands	  	& 8\\
Central frequency of each band (MHz)	& 139.3, 141.5, 143.7, 145.9,\\ 
					& 148.1, 150.3, 152.5, 154.7\\
Width of each band (MHz)		& 2.5\\
Frequency resolution (kHz)		& 4.9 (9.8 after tapering)\\
Time resolution (sec)			& 10\\
Angular resolution for the 3C196 field	& $2'\times 2' \rm{cosec}(\delta) \simeq 2' \times 2.7'$\\
Conversion factor for the 3C196 field	& 1~mJy~beam$^{-1}$ = 3.3~K\\
Angular resolution for the NCP field	& $2'\times 2' \rm{cosec}(\delta) \simeq 2' \times 2'$\\
Conversion factor for the NCP field	& 1~mJy~beam$^{-1}$ = 4.4~K\\
\hline
\end{tabular}
   \end{table}

Table~\ref{obs_table_3C196} summarizes the main characteristics of the observations. The eight spectral windows were chosen to provide a contiguous frequency coverage, with a small overlap between the bands. As happened for the observations of the Fan region (B09), large parts of the fourth spectral band, centred at 145.9~MHz, turned out to be too severely contaminated by radio interference to be useful, and the whole band was discarded from the analysis. After a Hanning taper was applied to the data, the channels were no longer independent. The even channels of every band were discarded and the remaining (odd) channels had a width of 9.8~kHz.

The data were reduced using the AIPS++\footnote{http://aips2.nrao.edu/docs/aips++.html} package. We integrated the standard AIPS++ distribution with routines which explicitly deal with flagging radio frequency interference (RFI) in low frequency data, determine the polarization calibration and perform the direction-dependent calibration.

\subsection{Determination of the flux scale}
\label{flux_scale}

3C196 is the standard flux calibration source used for low-frequency WSRT observations, therefore no other flux calibrator was targeted for the observations of that field. 3C196 is a very bright steep-spectrum radio galaxy with a smooth spectrum down to about 10 MHz (Laing \& Peacock 1980). We adopted a flux density at 150 MHz of 76.8~Jy for 3C196, and a power-law spectral index of $\alpha=0.64$ in the observed frequency range.

3C196 was also used as the primary flux calibrator for the observations of the NCP field. It was observed for about half an hour every night after the observation of the target field. The calibration solutions were transferred from 3C196 to the NCP field to set the flux scale. The System Equivalent Flux Density (SEFD) measured towards 3C196 is different from that measured towards the NCP, however, because 3C196 is located in one of the coldest spots of the 408~MHz map of the Galaxy (Haslam et al. 1982) and the Galactic emission significantly contributes to the system noise at low frequencies. 

The WSRT receivers operate with an automatic gain control system before the analog-to-digital converter which continuously measures the total power in order to allow corrections for the variable input levels. The total power detectors, which integrate the power over the whole 2.5~MHz band, are corrupted by RFI for most of the time, therefore the correlation coefficients cannot be corrected for the variations in the system noise. Therefore we corrected for the differences in the SEFD according to the following procedure, already applied in B09.

After the strongest RFI was removed from the data, we compared the power levels between the 3C196 and the NCP fields. In areas where the 2.5~MHz band was free of RFI we found that the ratio between the NCP and the 3C196 fields is $\sim$1.3 over the eight spectral bands. We applied this additional correction factor after the bandpass calibration was transferred from 3C196 to the NCP field.

\subsection{Calibration of the 3C196 field}
\label{calibration_3C196}

The calibration of the 3C196 field can be performed directly through the selfcalibration process, because at the frequency and angular resolution of our observations 3C196 is a point source bright enough to allow a signal-to-noise ratio (SNR) of $\sim$4 per baseline, per channel, per 10~sec. The calibration of the data proceeded as follows. 

The data were first flagged, in order to remove bright, time-variable and frequency-variable RFI. This was done by using flagging routines specifically developed to apply median filtering in the time and frequency domain to the visibility data. 

After that, selfcalibration solutions were computed both for amplitude gains and for phases at the highest time and frequency resolution available, in order to compensate for temporal and frequency variations of the gains induced by RFI or by ionospheric fluctuations. The initial sky model for the selfcalibration consisted of 3C196 alone. 

After the data were corrected, residual visibilities were formed by subtracting the model from the calibrated data. The
residual visibilities were binned and their distribution fitted with a Gaussian profile. All the data points further away
than five times the standard deviation of the Gaussian distribution were flagged. An image made with only the long baselines - $|{\bf u}| > 150$ - was created and deconvolved by using a Clark CLEAN algorithm down to a certain flux threshold. Point sources brighter than a defined flux threshold were extracted from the deconvolved map. The list of point sources served as a model for the following iteration of the selfcalibration. 

The process of creating a sky model, calibrating the data and flagging was repeated until no data were flagged and until no other sources could unambiguously be included in the the sky model. Four or five iterations were typically needed for this process to converge. 

\begin{figure}
\centering
\resizebox{1.0\hsize}{!}{\includegraphics[angle=-90]{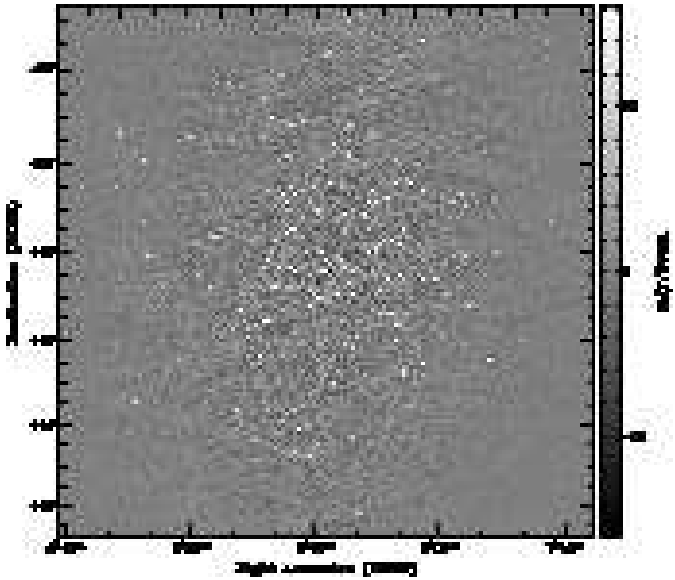}}
\caption{Stokes $I$ map of the 3C196 field made by using only the long baselines ($|{\bf u}| > 150$) and the data from the 139.3~MHz spectral band. All the point sources brighter than 150~mJy were extracted from this map and used as the final sky model for the selfcalibration of the band. The conversion factor is 1~mJy~beam$^{-1}$ = 3.3~K.}
\label{3C196_sky_model_spw1}
\end{figure}

Figure~\ref{3C196_sky_model_spw1} shows the sky map for the spectral band at 139.3~MHz made with only the long baselines obtained at the end of the selfcalibration loop. With that baseline selection, the sky appears to be dominated by point source emission. This justifies our choice of sky model. 

We note that the WSRT primary beam has a Half-Power Beam Width of about $6^\circ$ at 150~MHz, therefore a $12^\circ \times 12^\circ$ image maps the full primary beam. The images presented throughout the paper are not corrected for the primary beam attenuation unless stated otherwise. This implies that the noise across the image is constant but the sky brightness decreases when moving away from the centre because of the primary beam attenuation.
%
%

Table~\ref{selfcal_table_3C196} summarizes the selfcalibration model used for 3C196 in terms of detected sources and
flux limit. The slight difference in the flux threshold for source selection arises from the fact that some spectral
bands are more affected by RFI, therefore their calibration is less accurate than in other bands. We also note that
including CasA and CygA in the selfcalibration model makes no difference in the solutions. In our data CasA and CygA
are strongly suppressed by the primary beam and have an average peak flux in the 148.1~MHz band of $\sim$1.4~Jy and $\sim$1.8~Jy respectively.
\begin{table}	
\caption[]{The selfcalibration model for the 3C196 field.} 
\label{selfcal_table_3C196}
\begin{tabular}{l l l}        
Spectral band (MHz) & Flux threshold (mJy) & \# sources identified\\
\hline\hline 
139.3 & 150 & 140\\ 
141.5 & 150 & 140\\
143.7 & 250 & 102\\
148.1 & 180 & 128\\
150.3 & 250 & 83\\ 
152.5 & 150 & 140\\
154.7 & 100 & 217\\
\hline
\end{tabular}
   \end{table}

\begin{figure}
\centering
\resizebox{1.0\hsize}{!}{\includegraphics[angle=-90]{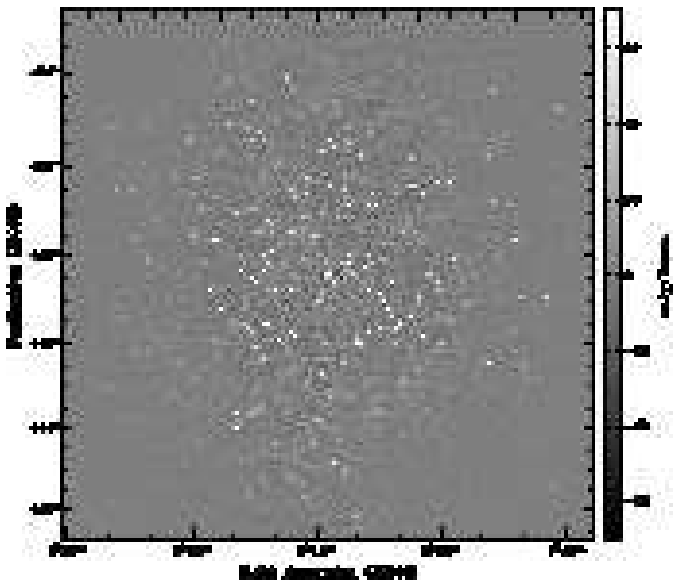}}
\caption{Stokes $I$ map of the 3C196 field made by averaging all the bands. The conversion factor is 1~mJy~beam$^{-1}$ = 3.3~K.}
\label{3C196_final}
\end{figure}

Figure~\ref{3C196_final} shows the average image from all the spectral bands after the selfcalibration process converged. We can already see that the image shows only point sources and that at this flux level, no diffuse emission associated with the Galaxy is visible.
%
%

In the field there are three sources with flux densities of $\sim$6~Jy: 3C197.1 located at $\alpha = 8^{\rm h}21^{\rm m}33.6^{\rm s}.0$, $\delta = 47^\circ 02' 37''$, 4C~+46.17 located at $\alpha = 8^{\rm h}14^{\rm m}30.3^{\rm s}.0$, $\delta = 45^\circ 56' 40''$ and 6C~B075752.1+501806 located at $\alpha = 8^{\rm h}01^{\rm m}35.4^{\rm s}.0$, $\delta = 50^\circ 09' 46''$. 

All these sources show peculiar spiky error patterns. These spiky patterns are due to ionospheric fluctuations which are different along different lines of sight. This effect is related to the one dimensional nature of the WSRT array, whose instantaneous response is a fan beam that rotates $180^\circ$ over 12~hours. Given this, spiky patterns are caused by phase variations in the telescope gains generated by temporal variations in the ionospheric electron column density. If ionospheric fluctuations vary across the field of view, they cause an effect on images which cannot be corrected with the traditional selfcalibration, which assumes a single phase correction over the entire field of view.

By inspecting the selfcalibration solutions we saw that the various nights showed quite different ionospheric behaviour. During the nights in November, the ionosphere turned out to be rather quiet throughout the whole observing run, whereas it started becoming more turbulent in the nights in December, with changes of tens of degrees of phase on time scales of minutes. This turbulence was non-isoplanatic, i.e. different lines of sight within the primary beam experienced different regions of turbulence. 

Since errors due to the ionospheric turbulence have a multiplicative effect, they are more prominent for the brightest sources. These spikes can reach up to 30-40~mJy~beam$^{-1}$.

In addition, 3C196 still shows phase errors due to imperfect calibration, and also calibration errors in the form of a ring-like pattern which surrounds the source itself. These errors can reach a brightness of 120-140~mJy~beam$^{-1}$.

In order to further improve the dynamic range of the image, we performed polarization calibration and looked for spurious residual signals in the XY and YX correlations.

3C196 was also used to calibrate the polarization leakages because it is unpolarized. Since the leakages are normally found to be rather stable with time for the WSRT, they were determined by computing solutions per $\sim$1~hour time interval. We checked that leakages remain stable over time by computing solutions every five minutes and we found no substantial difference.

After the leakage correction is applied (Sault, Hamaker \& Bregman 1996), a phase offset between the horizontal and the vertical dipoles still remains unknown and this causes the signal to leak from Stokes $U$ into Stokes $V$ (Sault, Hamaker \& Bregman 1996). In order to correct for this phase difference, the polarized pulsar PSRJ0218+4232 was observed for about 15~minutes just prior to the start of the 12~hour synthesis. Since the pulsar has a known rotation measure $RM = -61$~rad~m$^{-2}$ (Navarro et al. 1995), we corrected the phase difference by rotating the polarization vector in the plane defined by the Stokes $U-V$ parameters in order to have zero Stokes $V$ flux, where the direction of the rotation has to provide a negative RM for the pulsar.

After the calibration was performed, we inspected the XY and YX visibility data and we found the presence of noise a factor of $\sim$1.5 higher than the average, which remains correlated over periods of a few hours in the whole band. This increment affects only a few specific baselines which involve antennas 7 and 8, which are close to the WSRT control building. These baselines were not included in the images but were still included in the selfcalibration solutions. We suspect that faint RFI coming from the control building of the telescope may be responsible for the noise excess. We manually flagged those baselines at those specific time ranges and we recomputed the selfcalibration solutions.

The second way to improve the dynamic range was to correct for the effect of non-isoplanaticity in the ionosphere
towards the brightest sources. This can be done through the procedure known as
``peeling'': direction-dependent
calibration solutions are computed for each source that has to be peeled from the data (see, for instance, Mitchell et al. 2008 and Intema et al. 2009 as useful references for the peeling procedure). Peeling was implemented as follows. We subtracted our best estimate of the sky model from the visibilities apart from the source that we want to calibrate towards. We then solved for the complex antenna gains using only that specific source as sky model. In this way we account for time dependent position shifts due to ionospheric disturbance and possible time dependent amplitude errors due to small variations in the gains of the primary beams. The computed solutions were inverted and applied to the model of the source to be subtracted. In this way a ``corrupted'' model was created. This corrupted model was then subtracted from the visibilites.

The three sources that show strong ionospheric effects are not bright enough to be selfcalibrated per channel and per time slot. Therefore, we averaged all the channels within a single band and computed solutions every 10~sec.  

\begin{figure*}
\centering
\resizebox{0.44\hsize}{!}{\includegraphics[angle=-90]{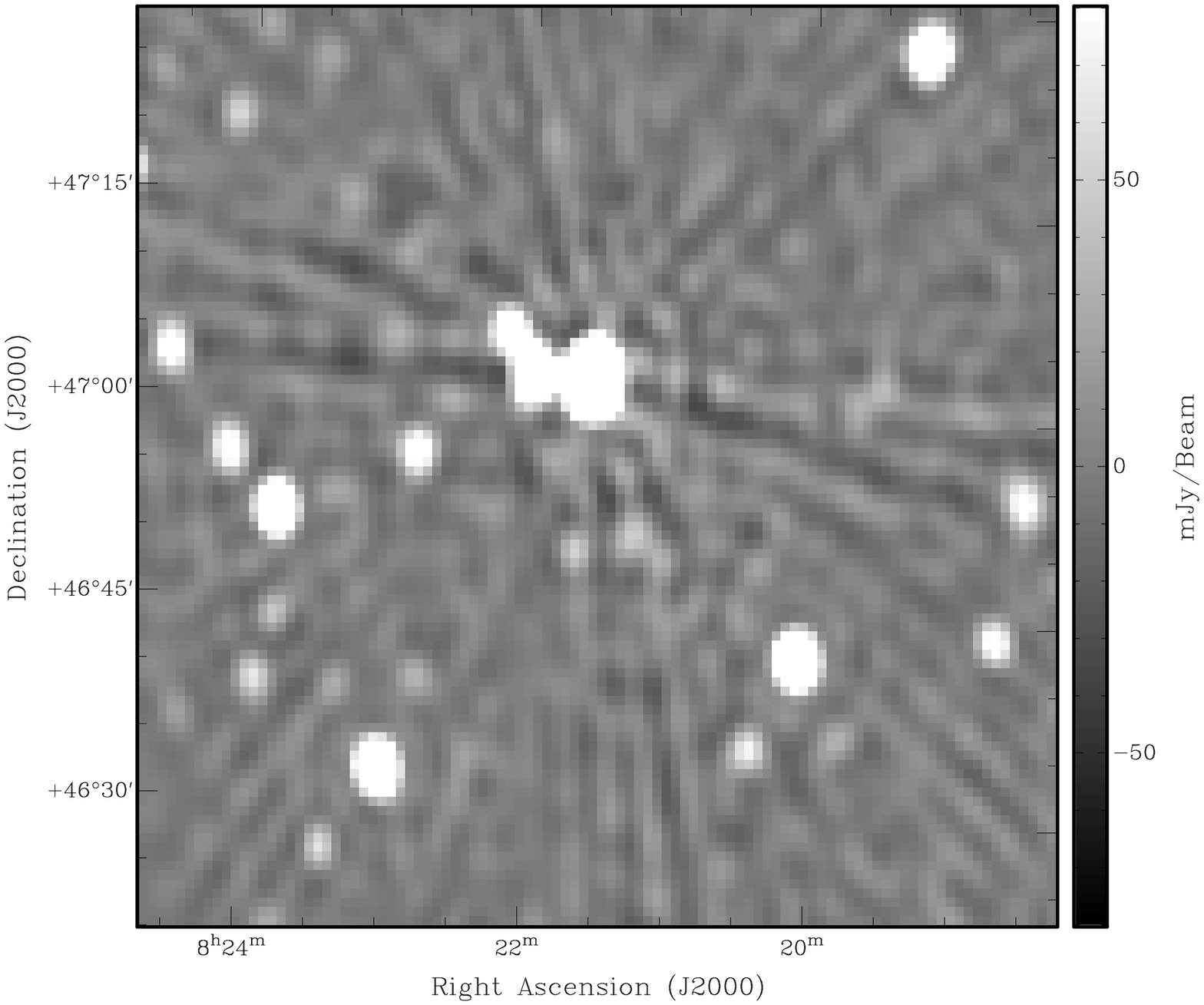}}
\resizebox{0.44\hsize}{!}{\includegraphics[angle=-90]{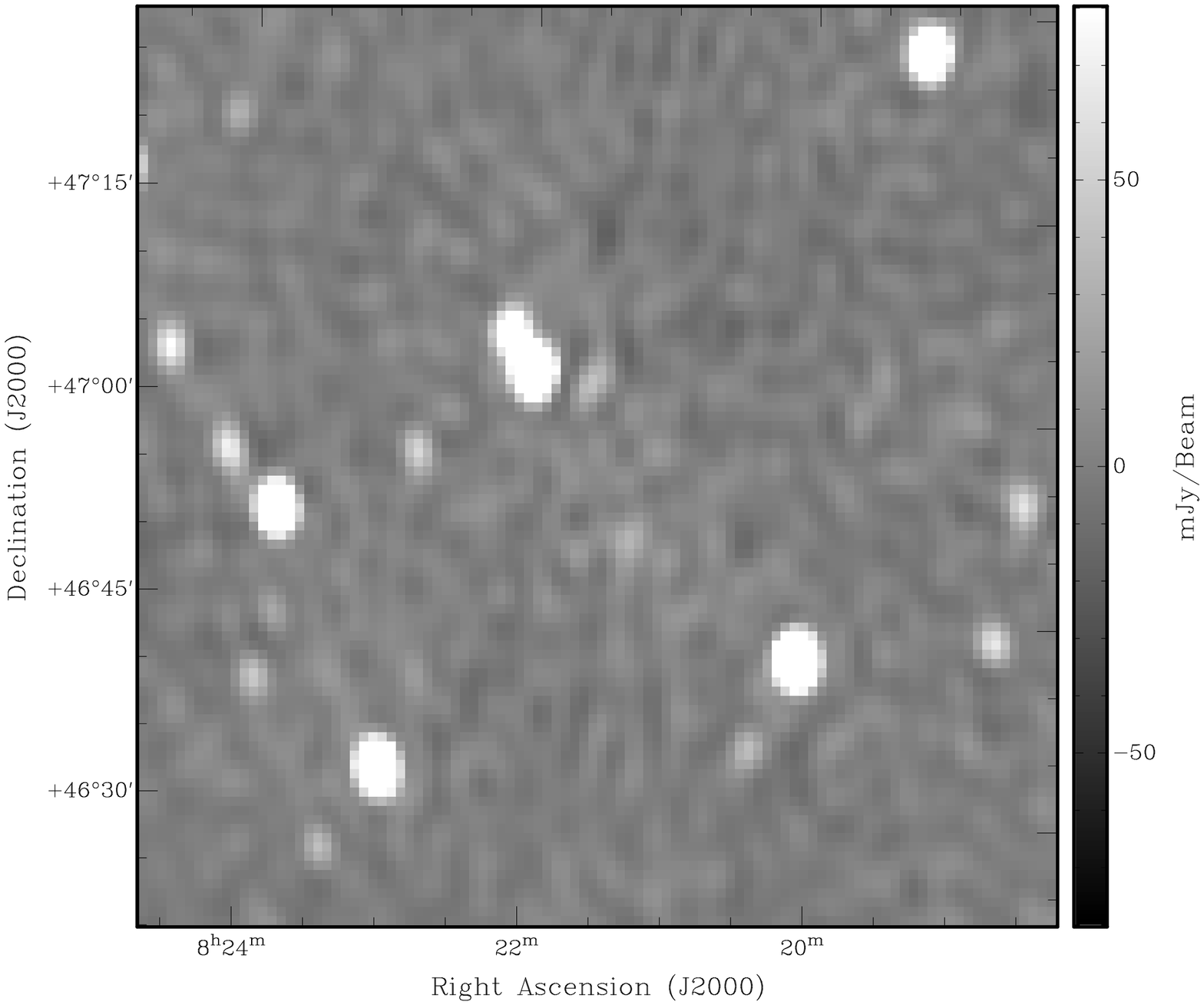}}
\resizebox{0.44\hsize}{!}{\includegraphics[angle=-90]{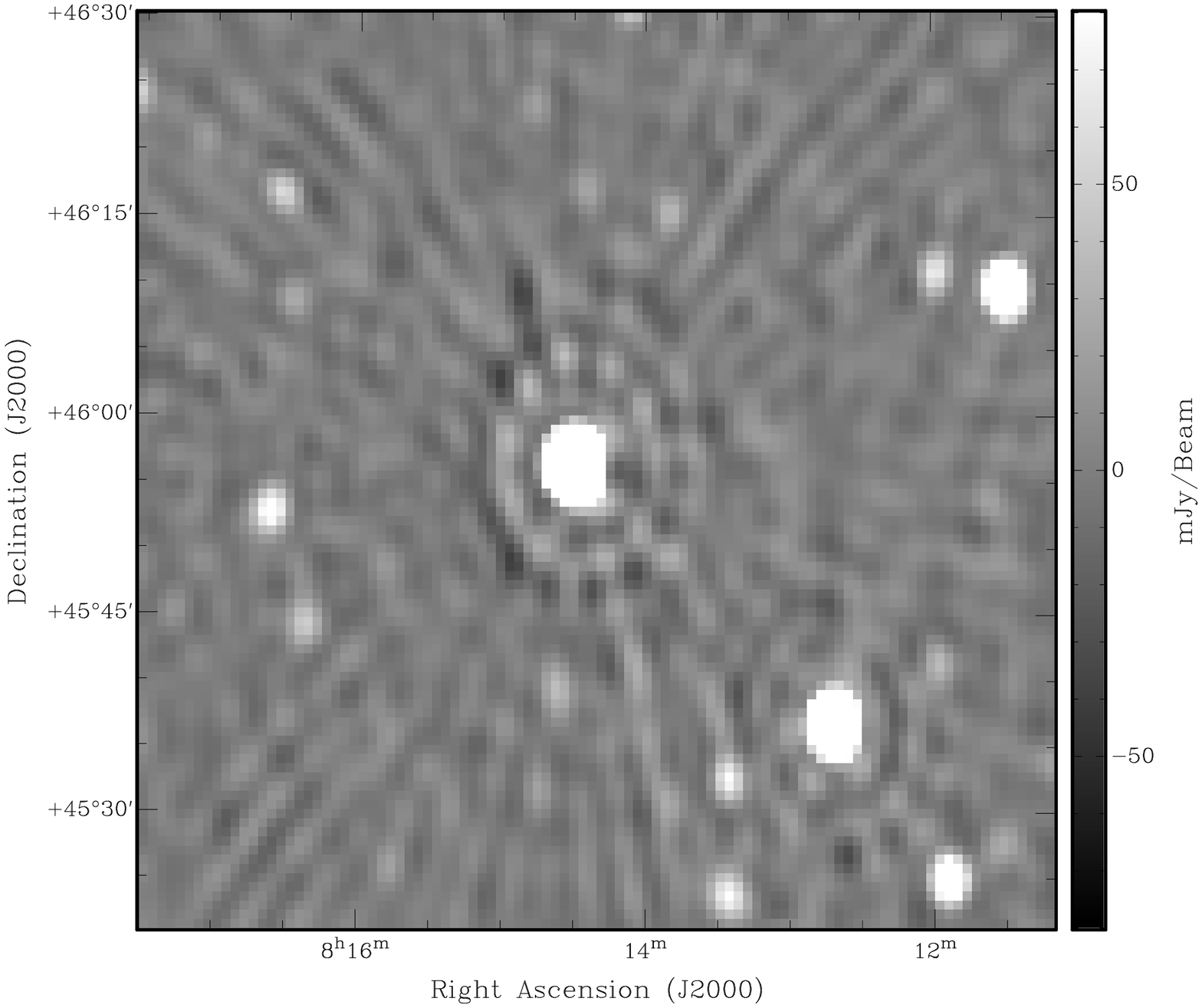}}
\resizebox{0.44\hsize}{!}{\includegraphics[angle=-90]{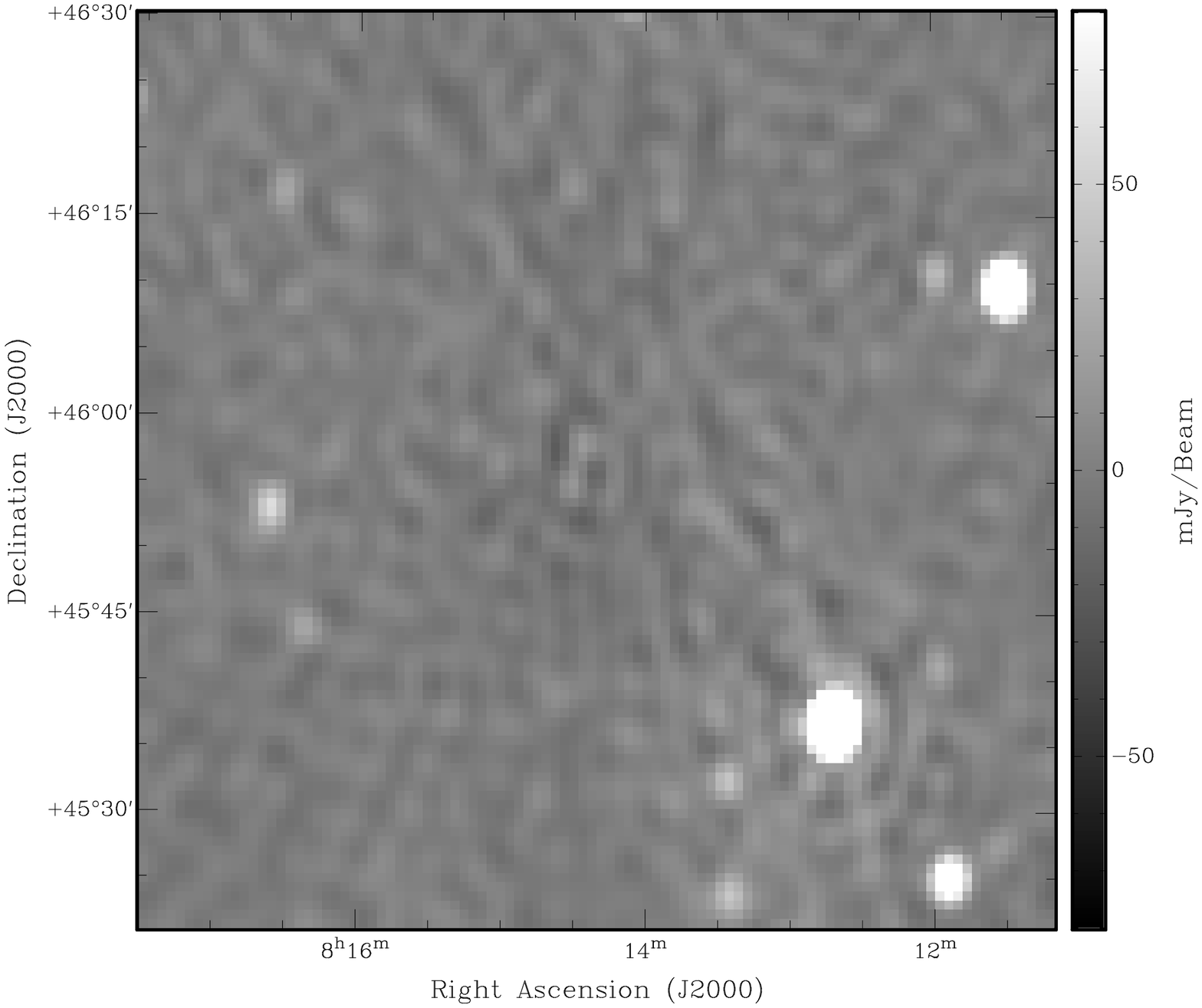}}
\resizebox{0.44\hsize}{!}{\includegraphics[angle=-90]{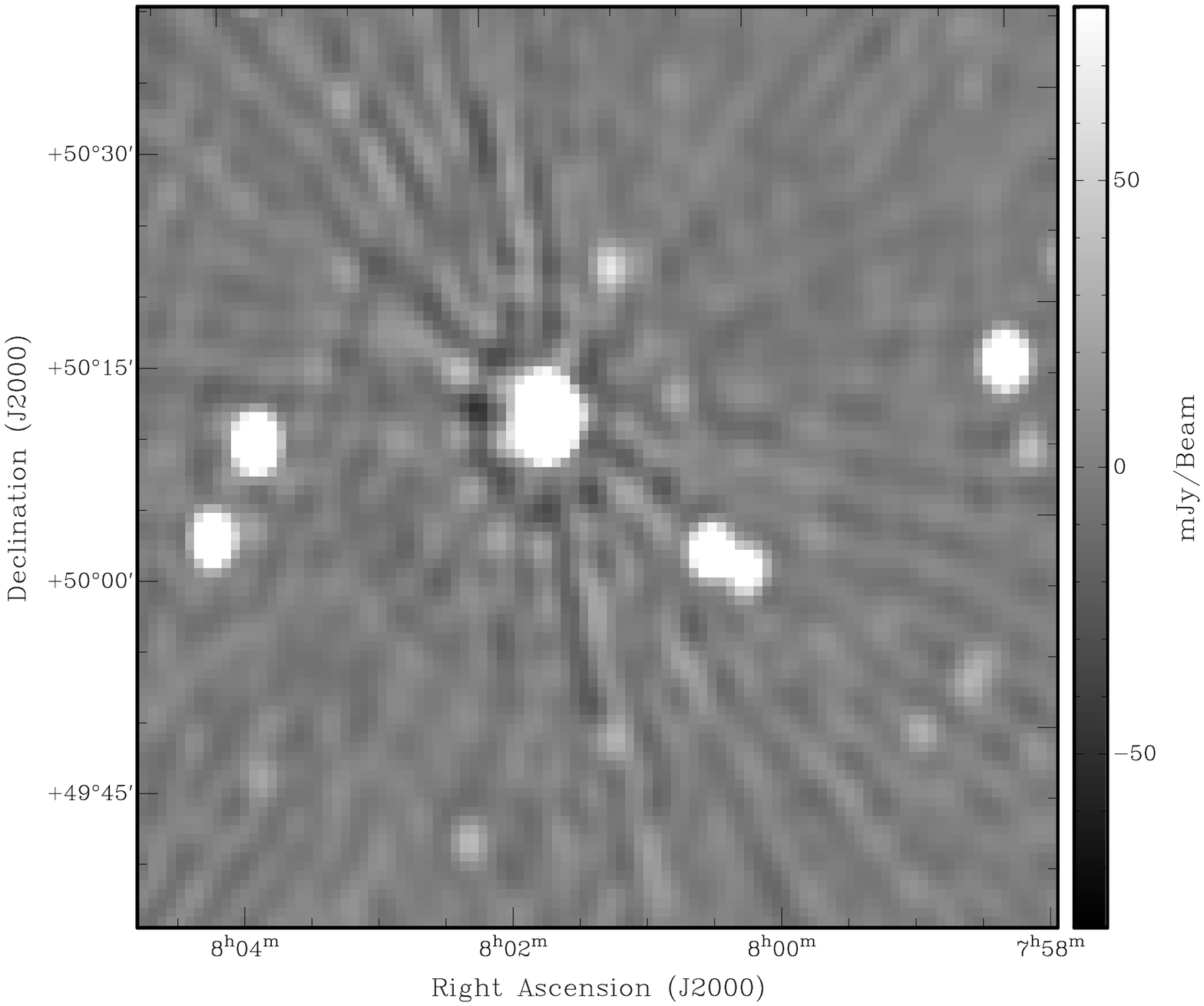}}
\resizebox{0.44\hsize}{!}{\includegraphics[angle=-90]{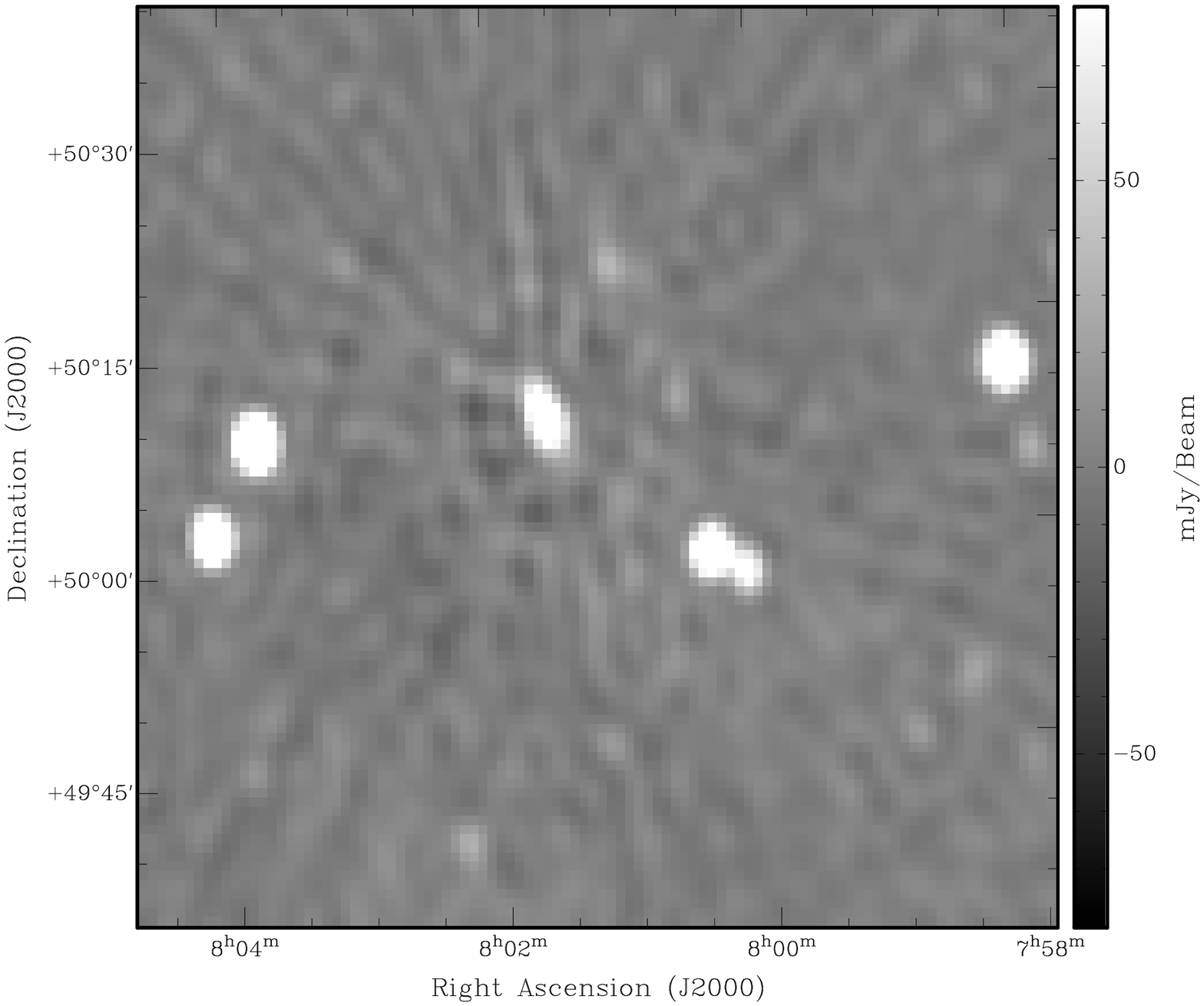}}
\caption{From top to bottom: the left column shows a zoomed-in image around
3C197.1, 4C~+46.17 and 6C~B075752.1+501806 respectively which highlights their
different spiky patterns due to non-isoplanaticity in the ionosphere. The right column shows the residual images after the sources have been peeled off. The images represent an average of the seven spectral bands of the 3C196 data.}
\label{3C196_peeled_sources}
\end{figure*}

Figure~\ref{3C196_peeled_sources} shows a comparison before and after peeling the sources. 
%
3C197.1 and 4C~+46.17 are very well subtracted, whereas 6C~B075752.1+501806 shows an evident residual which looks rather elongated in the North-East direction. The ratio between the $rms$ of the residual image around the position of the peeled source and the peak of the source itself gives a measure of the accuracy of the calibration for that source. We found that this ratio is $\sim$0.2\% for 3C197.1 and 4C~+46.17 and $\sim$4\% for 6C~B075752.1+501806. 

We found that the third source appears slightly extended, therefore modeling it with a
point source gives a low calibration accuracy. The residual is low enough not to affect the subsequent analysis and conclusions, however.

\begin{figure}
\centering
\resizebox{1.0\hsize}{!}{\includegraphics[angle=-90]{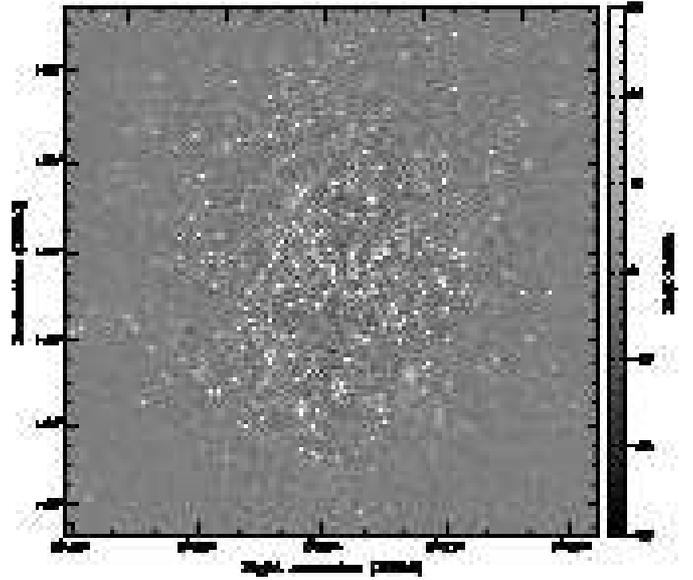}}
\caption{The 3C196 field after averaging all the spectral bands. Three sources were subtracted from the visibility data and a final selfcalibration loop was performed as described in the text. The conversion factor is 1~mJy~beam$^{-1}$ = 3.3~K.}
\label{3C196_final_image}
\end{figure}

After the data were manually flagged as described above and the three sources were
peeled off, the selfcalibration solutions were recomputed for each channel and for
every 30~sec in order to improve the SNR. The sky model was updated by excluding the
three peeled sources and the data were further flagged based upon the distribution of
the residual visibilities as described above. Cas A and Cyg A were also subtracted from the data. The resulting image is shown in Figure~\ref{3C196_final_image}.
%
%
3C196 now looks better calibrated than in Figure~\ref{3C196_final}. The rms value of the area around 3C196 where the error pattern was present, was indeed reduced by a factor of $\sim$6.

By generating a $20^\circ \times 20^\circ$ image we found that the observed noise at the edge of field is 0.5~mJy~beam$^{-1}$. This gives a dynamic range of $\sim$150000:1.

We note that the time ranges when the XY and YX correlations were affected by correlated noise influenced the selfcalibration solutions and were responsible for the errors around 3C196. Once those data were manually flagged and new selfcalibration solutions were computed and applied to the data, the error pattern disappeared. Only after this step was it possible to achieve a very high dynamic range.

We note, however, that we expect our data to be confusion limited rather than noise limited in the centre of the field. B09 found a confusion noise of $\sim$3~mJy~beam$^{-1}$ at 150~MHz and we expect a similar level of noise at the field centre for the 3C196 data as well. We will return to this point in Section~\ref{power_spec} where we analyse the power spectrum.

There are other features which become evident in the data in Figure~\ref{3C196_final_image}. Fainter sources now start to show their own spiky patterns due to the non-isoplanaticity of the ionosphere. We believe that the global radial pattern visible across the whole image is actually the result of all the ionospheric contributions related to each individual source.

Moreover, there is a mottled pattern spread over the whole image, consisting of
alternating negative and positive spots. The pattern is rather regular and the spots seem to have a characteristic size of 15-30~arcmin. They
also appear to be correlated with the position of the point sources: every
source seems to be surrounded by a positive ``pedestal''. 

B09 found faint diffuse Galactic emission in the Fan region down to
$\sim$10~arcmin scales. If compared with the fluctuations observed by B09, this
pattern looks much more regular. We found that this pattern is generated only by the baselines 7A and 8B, which were found to be corrupted, probably from broad-band faint RFI signals of local origin. In the further analysis we therefore discarded those baselines.

\subsection{Calibration of the NCP field}
\label{calibration_ncp}

The calibration of the NCP field followed closely the calibration of the 3C196 field. 

The data were initially flagged to remove time and frequency variable RFI.
Afterwards, the bandpass calibration was transferred from 3C196 and the
visibilities were corrected for the total power measurement as described in
Section~\ref{flux_scale}. At this point an initial sky model was generated by
making a sky map with only the long baselines,  $|{\bf u}| > 150$. The brightest
source in the field is the radio galaxy 3C061.1 which has an integrated flux of 31.2~Jy at 178~MHz
(Laing, Riley \& Longair, 1983). It has a size of approximately 10~arcmin along its major axis and is therefore resolved at our angular resolution. The sky model cannot be represented by point sources only, but has to be represented by a collection of CLEAN components.

The map made with only long baselines was deconvolved down to a certain flux
threshold by using a Clark CLEAN algorithm. The CLEAN components identified in a
$12^\circ \times 12^\circ$ image served as a model for the next iteration of the
selfcalibration. We stopped the deconvolution as soon as the first negative CLEAN component was found. The reason for this was that negative components might be a consequence of imperfect calibration, therefore they cannot be corrected for if they are included in the sky model used for selfcalibration. A sky model was created for each spectral window independently. 
The field does not contain enough flux to selfcalibrate the data per individual 9.8~kHz channel and every 10~sec as was done for 3C196. By averaging all the channels within a single band there was sufficient signal-to-noise ratio to compute solutions every 10~sec in order to correct for the most rapid fluctuations in the ionosphere. After the data were corrected, they were flagged based on the residual visibilities as described for the 3C196 data.

The process of creating a sky model, calibrating the data and flagging was
repeated until no data were flagged. Four or five iterations were
typically needed for the process to converge. Including CasA
and CygA in the selfcalibration model makes no difference in the solutions, as before. In
the NCP data CasA and CygA have an average peak flux in the 148.1~MHz band of $\sim$1~Jy and $\sim$3.6~Jy respectively.

\begin{figure}
\centering
\resizebox{1.0\hsize}{!}{\includegraphics[angle=-90]{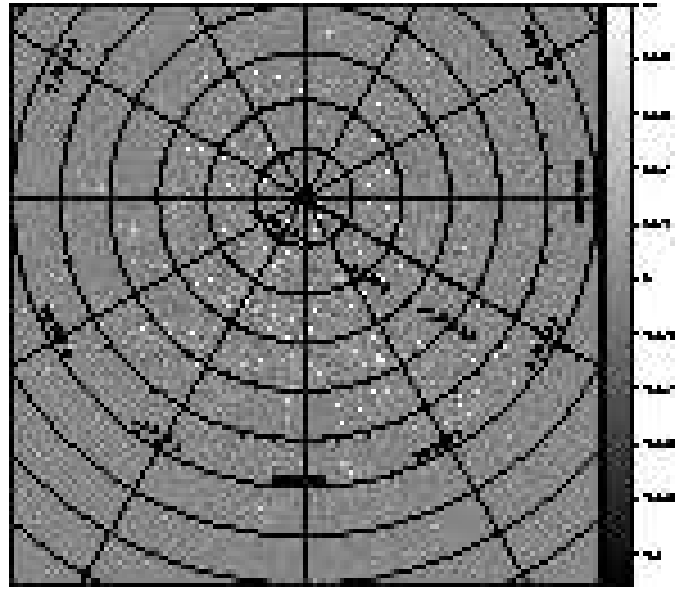}}
\caption{Stokes $I$ map of the NCP field made by using only the long baselines ($|{\bf u}| > 150$) and the data from the 139.3~MHz spectral band. All the CLEAN components brighter than 110~mJy detected in this map were used as a final sky model for the selfcalibration of the band. The units are Jy~beam$^{-1}$ and the conversion factor is 1~mJy~beam$^{-1}$ = 4.4~K.}
\label{ncp_high_res}
\end{figure}

Figure~\ref{ncp_high_res} shows the image made with only the long baselines. This image was used to extract the best sky model for the spectral band at 139.3~MHz.
%
%

\begin{figure}
\centering
\resizebox{1.0\hsize}{!}{\includegraphics[angle=-90]{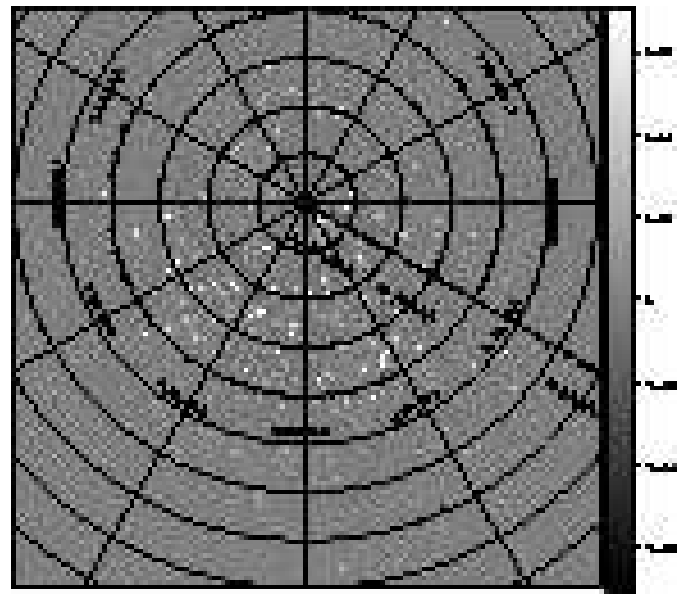}}
\caption{Stokes $I$ map of the NCP field made by averaging all the spectral bands after selfcalibration. The units are Jy~beam$^{-1}$ and the conversion factor is 1~mJy~beam$^{-1}$ = 4.4~K.}
\label{ncp_final}
\end{figure}

Figure~\ref{ncp_final} shows the selfcalibrated image.
%
%
There are a few similarities between the selfcalibrated map of the NCP field and the 3C196 image of Figure~\ref{3C196_final}. In the NCP field no diffuse emission appears to be evident in the map, as for 3C196, and there are three sources which have spiky patterns similar to those associated with the bright sources of the 3C196 field. In this case they appear as bright negative values that can rise up to $\sim$40~mJy~beam$^{-1}$.

Comparing with existing catalogues, these sources have NRAO VLA Sky Survey (NVSS, Condon et al. 1998) sources as their closest counterparts
and are not present in any of the 3C catalogues. The brightest
source after 3C61.1 is $\sim$5.8~Jy and can be associated with an NVSS source
located at $\alpha = 1^{\rm h}17^{\rm m}32.8^{\rm s}.0$, $\delta = 89^\circ 28'
49''$. The second brightest source is $\sim$4.1~Jy and can be associated with an
NVSS source located at $\alpha = 1^{\rm h}09^{\rm m}17.6^{\rm s}.0$, $\delta =
87^\circ 41' 18''$. The third brightest source is $\sim$3.4~Jy and can be
associated with an NVSS source located at $\alpha = 6^{\rm h}22^{\rm m}05.5^{\rm s}.0$, $\delta = 87^\circ 19' 49''$.

In order to detect possible faint, diffuse Galactic emission it is necessary to correct for errors in the direction of these three sources. We computed direction-dependent calibration towards those sources in the same way that we did for the sources in 3C196 data. The only difference was that in this case the sources are fainter, therefore the visibilities were averaged over all the channels in a single band and over 1~min in order to achieve a sufficient SNR.

The comparison between the maps before and after source peeling shows results similar to what was presented in Section~\ref{calibration_3C196}. We quantified the accuracy of the calibration towards each source by computing the ratio between the $rms$ of the residual after the source was subtracted and the peak flux, as we did for the sources in the 3C196 field. We found that the sources are subtracted with an accuracy of 0.4\%, 0.8\% and 0.8\% respectively from the brightest source to the faintest one.

\begin{figure}
\centering
\resizebox{1.0\hsize}{!}{\includegraphics[angle=-90]{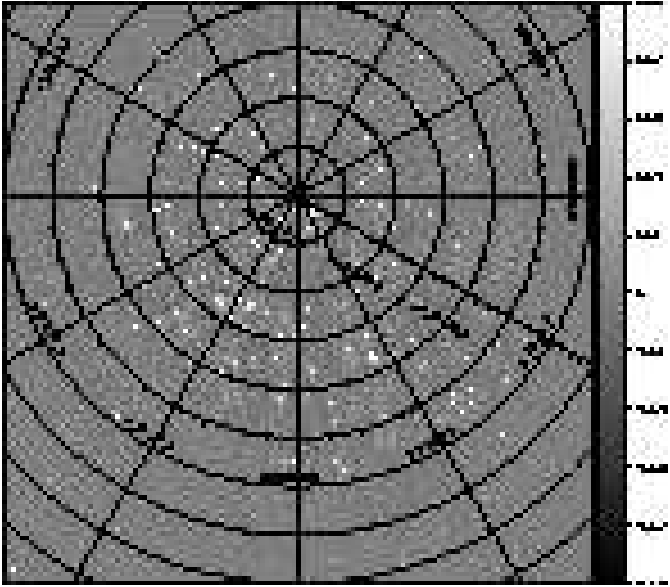}}
\caption{Stokes $I$ map of the NCP field made by averaging all the spectral bands after source peeling. The units are Jy~beam$^{-1}$ and the conversion factor is 1~mJy~beam$^{-1}$ = 4.4~K.}
\label{ncp_final_peeled}
\end{figure}

Figure~\ref{ncp_final_peeled} shows the image of the NCP field after the sources were peeled off.
%
%
Weak residuals are still visible around the peeled sources, and around other ones mainly located at the edge of the field. These residuals are again due to the ionospheric contribution. By generating a $20^\circ \times 20^\circ$ image we found that the observed noise at the edge of field is $\sim$0.7~mJy~beam$^{-1}$. 

In Figure~\ref{ncp_final_peeled} a mottled pattern spread over the whole image is also visible. It is similar to the one present in the final map of 3C196 but fainter, with a characteristic size $\sim$15-30~arcmin. We found that its origin is again due to the corrupted baselines 7A and 8B, which were therefore discarded from the following analysis.

The calibration of the XY and YX correlations was performed similarly to the 3C196 field. The leakages were computed using 3C196 which was observed for half an hour every run. The solutions were applied to the NCP target. The remaining unknown offset between Stokes $U$ and $V$ was then calibrated, again using the pulsar PSRJ0218+4232, which was observed for $\sim$20~min just before the NCP target.

\section{Polarization analysis}
\label{pol_an}

We applied the RM synthesis technique (Brentjens \& de Bruyn 2005) to analyse the polarized emission for both the 3C196 and the NCP field. Since the frequency setup is the same as the observations of the Fan region (B09), we briefly summarize here the most relevant points of the RM synthesis analysis.

The RM synthesis technique takes advantage of the Fourier relationship which exists
between $P(\lambda^2)$ and $F(\phi)$:
\begin{eqnarray}
    P(\lambda^2) = W(\lambda^2) \int^{+\infty}_{-\infty} F(\phi) \, \rm{e}^{2i\phi\lambda^2} \rm{d} \phi
\label{RM_synth_eq}
\end{eqnarray}
where $P(\lambda^2)$ is the complex polarized surface brightness, $W(\lambda^2)$ is a weighting function, $F(\phi)$ is the complex polarized surface brightness per unit of Faraday depth, $\lambda$ is the observing wavelength and $\phi$ is the Faraday depth. The output of the RM synthesis analysis is a cube of polarized maps at selected values of Faraday depth. The Fourier transform of  $W(\lambda^2)$ gives the RM spread function (RMSF), which is the resolution in Faraday depth.

Table~\ref{RM_table} summarizes the most relevant parameters for the polarization analysis. The first three parameters reported in  Table~\ref{RM_table} are defined by the relationships (61), (62) and (63), of Brentjens \& de Bruyn (2005): $\delta \phi$ is the resolution in Faraday depth, $\phi^{\rm scale}_{\rm max}$ is the maximum Faraday depth scale to which our observations are sensitive and $||\phi_{\rm max}||$ is the maximum RM value measurable.

It is worth noting that our observations are only sensitive to Faraday depths
smaller than the RMSF width, so that we are sensitive only to Faraday thin regions, with an extension in RM space of $\sim$1~rad~m$^{-2}$ or less. 
\begin{table}	
\caption[]{Summary of RM synthesis cube parameters} 
\label{RM_table}
\centering
\begin{tabular}{l l}        
\hline\hline \noalign{\smallskip}
   
$\delta \phi$				& 3.4~rad~m$^{-2}$\\
$\phi^{\rm scale}_{\rm max}$		& 0.85~rad~m$^{-2}$\\
$||\phi_{\rm max}||$			& 2650~rad~m$^{-2}$\\ 
Angular resolution for the 3C196 data	& 4.3~arcmin\\
Conversion factor for the 3C196 data	& 1~mJy~beam$^{-1}$ = 0.94~K\\
Angular resolution for the NCP data	& 4.2~arcmin\\
Conversion factor for the NCP data	& 1~mJy~beam$^{-1}$ = 1.3~K\\
\hline
\end{tabular}
\end{table}

Both fields were analysed as follows. As explained in Section~\ref{obs_res}, we
corrected for the on-axis instrumental polarization, but the WSRT has a strong
off-axis instrumental polarized response which increases with distance from the
centre of the image and can reach up to 20-30\% at the beam half-power radius.
The off-axis polarization is strongly frequency dependent and the polarization beam pattern is not well known at these frequencies. 

Since we are interested in detecting diffuse polarized emission coming from the Galaxy, we want to subtract most of the point sources which are likely to be instrumentally polarized, expecially those ones located at the edge of the primary beam.

Therefore, we constructed a sky model by making Stokes $I$, $Q$, $U$ and $V$ images for each spectral band by using only the baselines with $|{\bf u}| > 150$, as we did to determine the sky model for the selfcalibration. These images were deconvolved through a CLEAN deconvolution and all the sources down to 50~mJy were identified and removed from the data. CasA and CygA were also modeled in all the Stokes parameters for each spectral band and subtracted from the data.

Since most of the Galactic polarized emission appears on spatial scales greater
than a few arcmin, after the instrumentally polarized sources were removed from the data we applied a taper in the $uv$ plane to lower the angular resolution down to $\sim$4~arcmin, and made a residual image for each channel. 

We computed a histogram distribution of the rms for all the Stokes $Q$ channel images and fit it with a Gaussian profile. Images which had a Stokes $Q$ rms value greater than three times the standard deviation of the Gaussian distribution were discarded from the RM synthesis analysis. Approximately 1500 channel images were used as input for the RM synthesis.

For the output RM cube, we selected the RM interval $[-50,50]$~rad~m$^{-2}$, where the Galactic emission most likely appears. In the following sections we describe the polarization results for the 3C196 and NCP data separately.

\subsection{Polarized emission in the 3C196 area}
\label{pol_3C196}

\begin{figure}
\centering
\resizebox{1.0\hsize}{!}{\includegraphics[angle=-90]{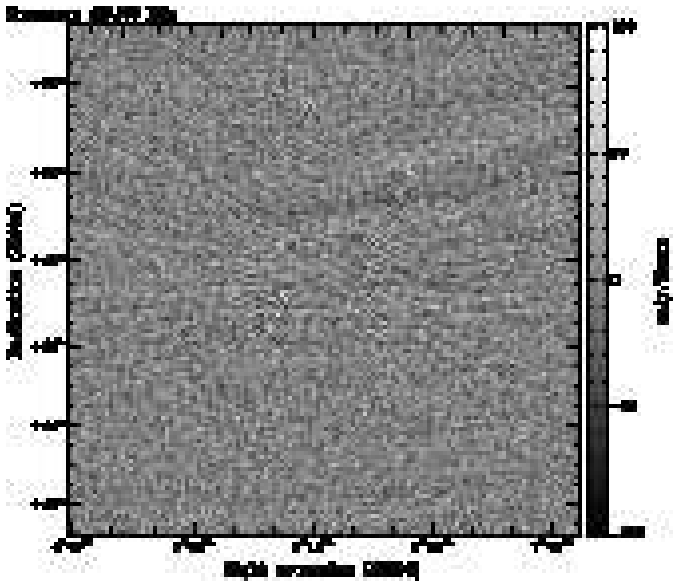}}
\caption{Stokes $U$ map of one of the channels of the spectral band at 139.3~MHz of the 3C196 data. The``whiskers'' are visible around the centre of the image. The angular resolution here is 4.3~arcmin and the conversion factor is 1~mJy~beam$^{-1}$ = 0.94~K.}
\label{whiskers_ex}
\end{figure}

In the individual channel maps of Stokes $U$, we have observed instrumental artifacts with a peculiar shape which make them look like ``whiskers''. As can be seen from Figure~\ref{whiskers_ex}, they look like alternating positive and negative stripes around the centre of the image. They appear in almost all the individual channel maps of all the spectral bands with an intensity which rises up to 70-80~mJy~beam$^{-1}$. They are not constant over the 12~hour synthesis and they do not repeat themselves at the same position for different frequencies. They appear in Stokes $U$ and $V$ but not at all in Stokes $Q$ and $I$. It is worth noting that Figure~\ref{whiskers_ex} shows no evident residual emission either from diffuse structures or from point sources.
%
%

The origin of such an artifact - of clear instrumental origin - is still
unknown. From the spatial location it seems to be somewhat associated with 3C196; de Bruyn \& Brentjens (2005) also find ``whiskers'' associated with bright sources in observations of the Perseus cluster at 350~MHz. They could not identify the exact cause of the artifact either.

\begin{figure*}
\centering
\resizebox{0.45\hsize}{!}{\includegraphics[angle=-90]{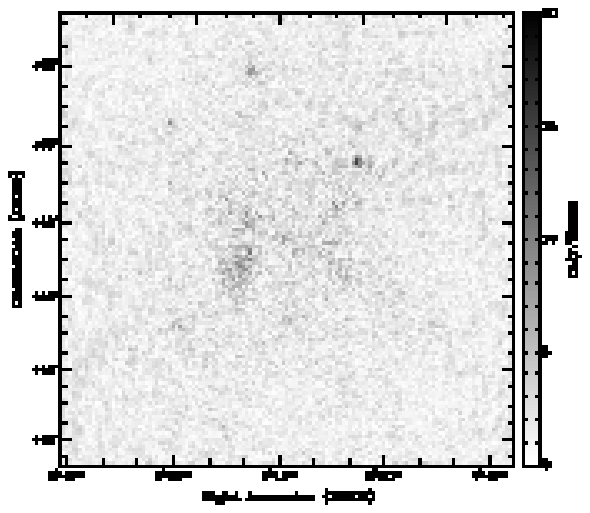}}
\resizebox{0.45\hsize}{!}{\includegraphics[angle=-90]{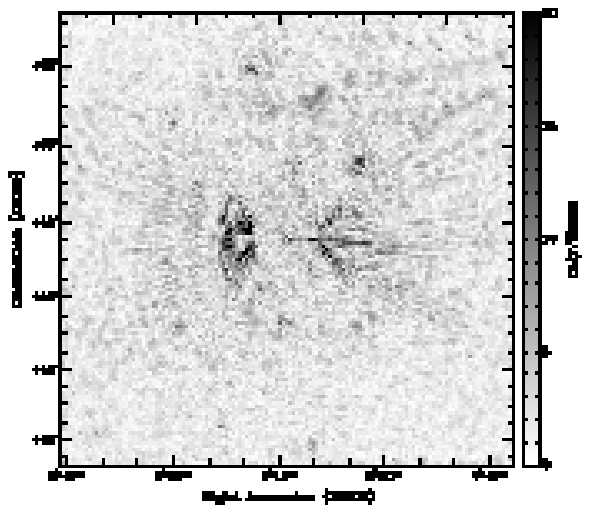}}
\resizebox{0.45\hsize}{!}{\includegraphics[angle=-90]{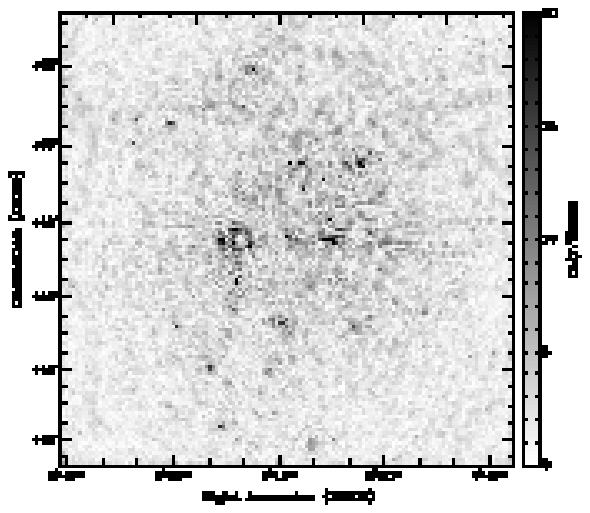}}
\resizebox{0.45\hsize}{!}{\includegraphics[angle=-90]{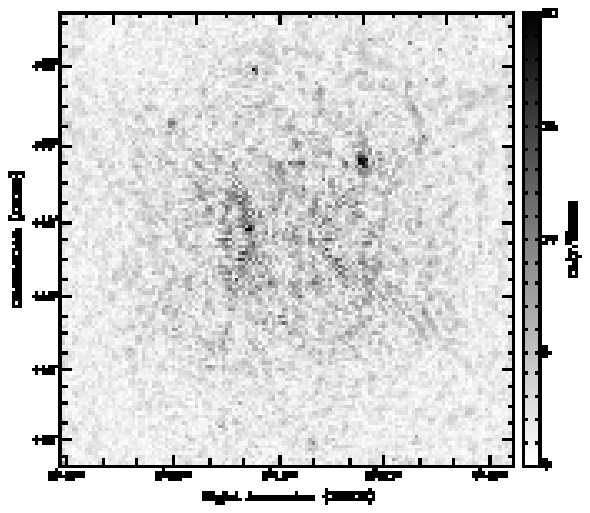}}
\caption{Polarized intensity of four frames of the RM cube in the 3C196 field:
-8~rad~m$^{-2}$ (top left), -2~rad~m$^{-2}$ (top right), 0~rad~m$^{-2}$ (bottom
left), 4~rad~m$^{-2}$ (bottom right). The sidelobes emanating from 3C196 are
clearly visible. The angular resolution here is 4.3~arcmin and the conversion factor is 1~mJy~beam$^{-1}$ = 0.94~K.}
\label{3C196_rm_cube}
\end{figure*}

In our case, the fact that these whiskers are variable with time, frequency and position in the image becomes a severe problem for the RM synthesis because these artificial signals affect many Faraday depth frames of the RM cube. Figure~\ref{3C196_rm_cube} shows a few frames of the output from the RM synthesis selected in the most relevant range of RM values.
%
%

The RM cube shows no polarized emission at Faraday depths greater than
$|10|$~rad~m$^{-2}$. At Faraday depths between -10~rad~m$^{-2}$ and
10~rad~m$^{-2}$ the polarized emission is highly contaminated by the
whiskers which are by far the brightest feature in the cube. Since they cannot be
removed from the data, it is not viable to separate the possible true emission of the sky and the instrumental one.

Since the whiskers are present only in Stokes $U$, however, it is possible to
perform an RM synthesis with only Stokes $Q$ in order to avoid their effect. An
RM synthesis with Stokes $Q$ alone implies that the RM cube will be symmetric
with respect to zero because the information about the sign of the RM is lost. Moreover, the noise is increased by a factor $\sqrt{2}$ while the same amount is lost in the signal if Stokes $Q$ and $U$ have statistically the same power.

\begin{figure*}
\centering
\resizebox{0.45\hsize}{!}{\includegraphics[angle=-90]{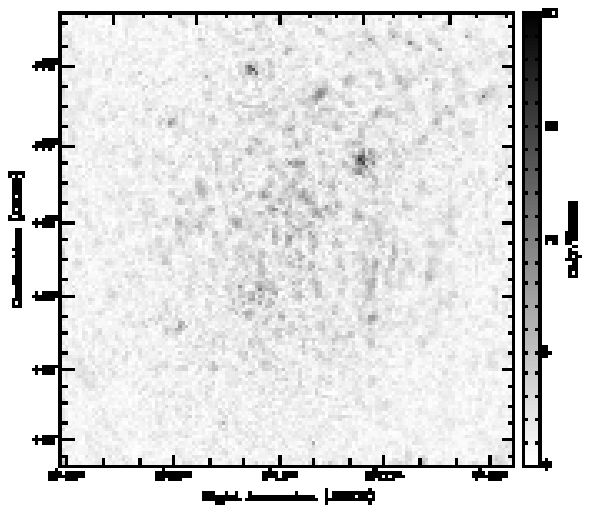}}
\resizebox{0.45\hsize}{!}{\includegraphics[angle=-90]{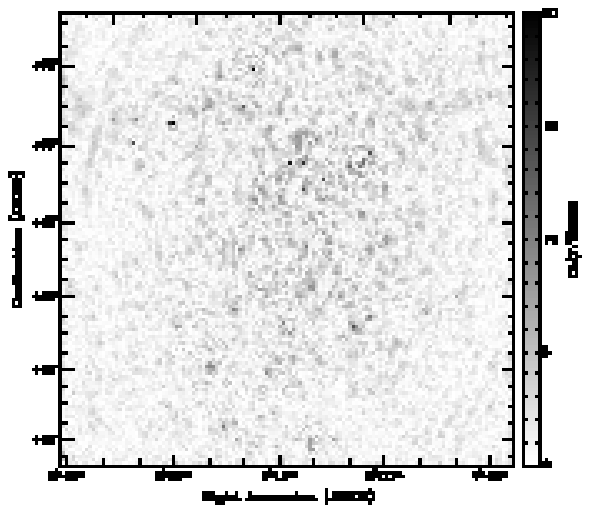}}
\caption{Polarized intensity of two frames of the RM cube made with Stokes $Q$ alone in the 3C196 field: -3~rad~m$^{-2}$ (left) and 0~rad~m$^{-2}$ (right). The angular resolution here is 4.3~arcmin and the conversion factor is 1~mJy~beam$^{-1}$ = 0.94~K.}
\label{3C196_rm_cube_only_Q}
\end{figure*}

The two most prominent frames of the RM cube made with Stokes $Q$ alone are
displayed in Figure~\ref{3C196_rm_cube_only_Q}.
%
%
Patchy diffuse polarized emission is observed throughout the whole field and is due to the corrupted baselines 7A and 8B. As before, they were therefore removed from the further analysis. A few instrumentally polarized point sources are still visible but we note that there is no residual associated with 3C196 in the map, indicating that the Stokes $Q$ parameter could be calibrated down to the thermal noise.

The RM cube determined through Equation~\ref{RM_synth_eq} still suffers from the sidelobe structure of the RMSF, caused by the incomplete $\lambda^2$ coverage of our observations. This problem can be alleviated by a
deconvolution from the RMSF through a CLEAN technique, analogous to the traditional two-dimensional
CLEAN of synthesis images. We used a publicly available
code\footnote{http://www.astron.nl/$\sim$heald/software} which performs the analogue of the Hogbom
CLEAN of the RM cube. We briefly describe the procedure here but refer to the more detailed
description contained in Heald, Braun \& Edmonds (2009).

The deconvolution is performed in the complex domain by identifying peaks in the spectrum obtained by cross-correlating the complex Stokes $Q(\phi)$ and $U(\phi)$ and the complex RMSF. At the location of each peak thus found, another search for a peak in the polarized intensity $P(\phi)$ above a certain threshold is performed. We chose three times the noise for the threshold. If a peak is found, a shifted and scaled complex version of the RMSF is subtracted from the Stokes $Q(\phi)$ and $U(\phi)$ at that pixel. The scaled version of the RMSF represents a CLEAN component. This loop was repeated until no peaks were found.

\begin{figure*}
\centering
\resizebox{0.45\hsize}{!}{\includegraphics[angle=-90]{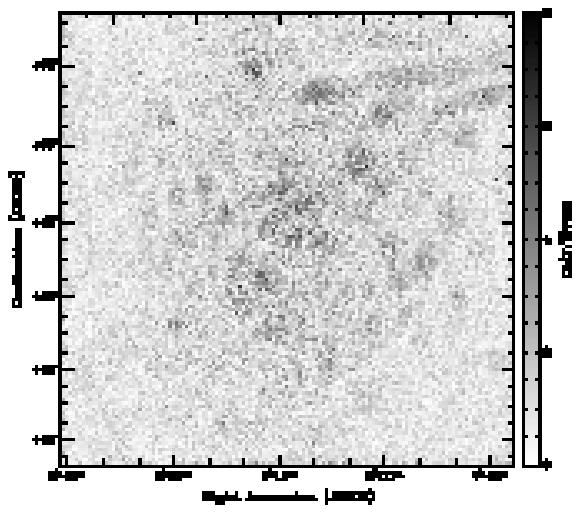}}
\resizebox{0.45\hsize}{!}{\includegraphics[angle=-90]{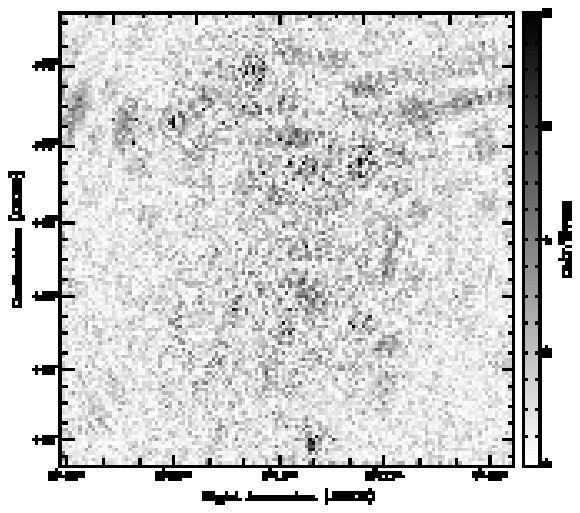}}
\caption{Polarized intensity of the two brightest frames of the RM cube in the 3C196 field: -3~rad~m$^{-2}$ (left) and 0~rad~m$^{-2}$ (right). The RM cube was made with Stokes $Q$ alone, without baselines 7A and 8B and was deconvolved from the RMSF. The angular resolution here is 4.3~arcmin and the conversion factor is 1~mJy~beam$^{-1}$ = 0.94~K.}
\label{3C196_rm_cube_clean}
\end{figure*}

The CLEAN components were then convolved with a Gaussian beam of full width at half maximum (FWHM)
equal to the FWHM of the RMSF and added to the residual cubes. In this way, restored $P$, $Q$ and
$U$ RM cubes were formed. Figure~\ref{3C196_rm_cube_clean} displays the two frames of the restored RM cube already shown in Figure~\ref{3C196_rm_cube_only_Q}. 
%
%

The frames show several patchy polarized structures across the whole field of view. In the
North-West corner of the map several stripes originating from time-variable RFI are visible. The frame
at RM~=~0 shows a few instrumentally polarized point sources.

For the power spectrum analysis we made an image of the total polarized emission $P$ (Brentjens 2007; B09):
\begin{eqnarray}
    P = B^{-1} \sum_{i=-7}^0 \left( P_{i} - \sigma_P \sqrt{\frac{\pi}{2}} \right)
\end{eqnarray}
where $\sigma_P$ is the noise in polarization, $P_i$ is the polarization map at the RM value $i$ and the sum is over the frames of the RM cube which show emission. The factor $B$ represents the area of the restoring beam divided by the interval between two frames of the RM cube. 

\begin{figure}
\centering
\resizebox{1.0\hsize}{!}{\includegraphics[angle=-90]{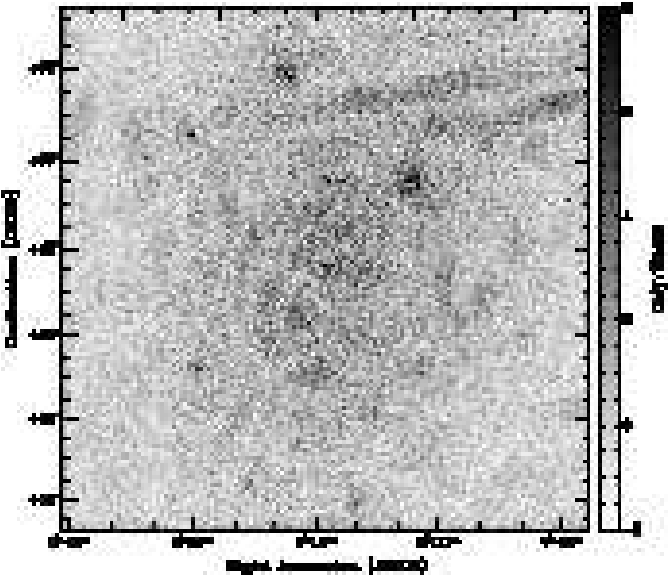}}
\caption{Map of the total polarized emission in the 3C196 area after integrating along the Faraday depth. The angular resolution here is 4.3~arcmin and the conversion factor is 1~mJy~beam$^{-1}$ = 0.94~K.}
\label{3C916_total_polarization}
\end{figure}

Figure~\ref{3C916_total_polarization} displays the integrated polarized intensity. It weakens moving away
from the image centre because of the primary beam attenuation, and this indicates that the detected
emission is true emission from the sky. Moreover, the fact that the polarized emission has no clear
counterpart in total intensity and that it varies with Faraday depth further proves that it is true
emission from the sky, probably coming through the Faraday screen mechanism (Wieringa et al. 1993; Gaensler et al. 2001; Bernardi et al. 2003; 
Wolleben \& Reich 2004; Schnitzeler 2007). 
%
%

\subsection{Polarized emission around the NCP}

The detection of polarized emission in the NCP area was hampered by several instrumental problems too. In all the individual channel images we observed broad arc-like features centred around the NCP. These features are not persistent over the whole 12~hour synthesis and are highly variable in time and frequency. They clearly do not belong to the sky.

The most plausible explanation for these artifacts is related to the fact that the NCP is the location of a constant geometrical delay: all the signals which show no difference in the geometrical delay add up at the NCP. In the case of the WSRT, these signals are mainly RFI coming from the telescope control building or the nearby surroundings.

These RFI signals are visible in all the Stokes parameters but they start to appear in the total intensity maps only when most of the sources are subtracted. The reason for this is mainly their intrinsic weakness, but also the fact that they are highly variable with time and frequency and, therefore, their intensity is averaged out over the 12~hour synthesis and over the whole spectral band.

Since the polarized signals are much fainter than the total intensity emission, they are the most prominent features in the Stokes $Q$ and $U$ maps.

\begin{figure*}
\centering
\resizebox{0.45\hsize}{!}{\includegraphics[angle=-90]{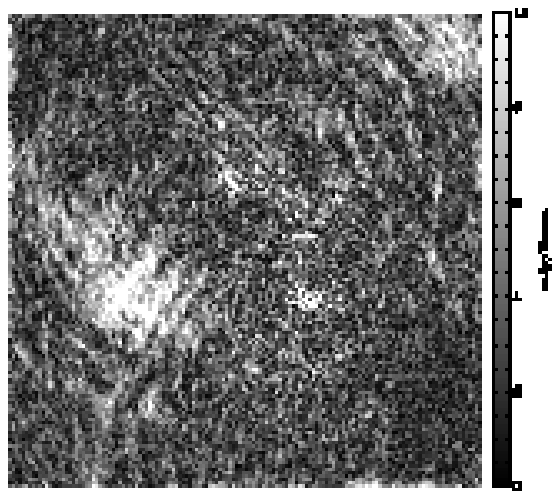}}
\resizebox{0.45\hsize}{!}{\includegraphics[angle=-90]{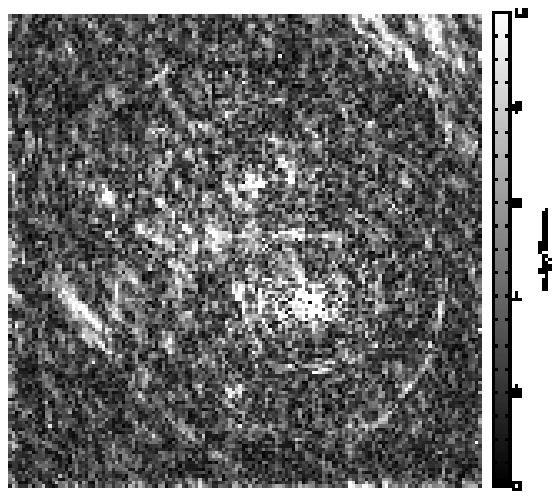}}
\resizebox{0.45\hsize}{!}{\includegraphics[angle=-90]{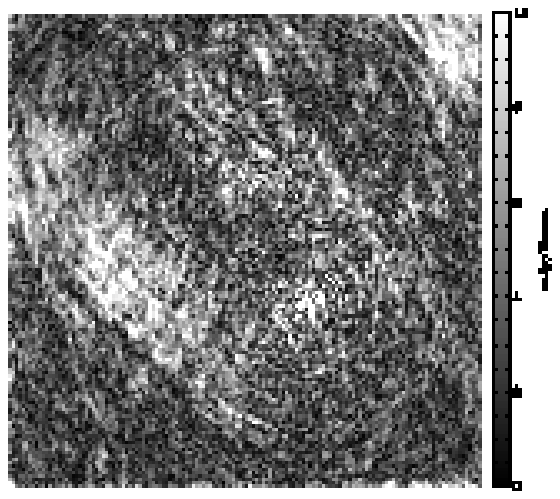}}
\resizebox{0.45\hsize}{!}{\includegraphics[angle=-90]{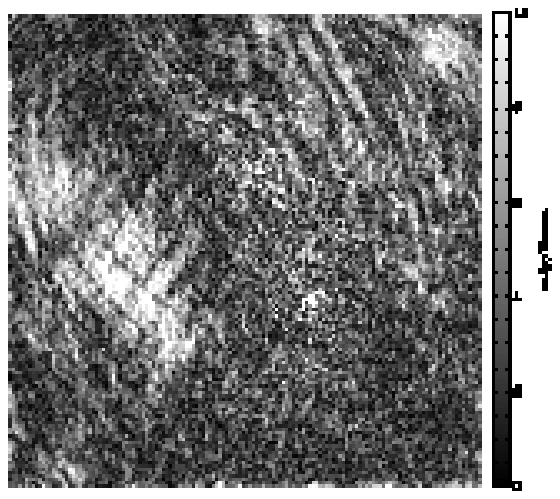}}
\caption{Polarized intensity of four frames of the RM cube in the NCP field: -12~rad~m$^{-2}$ (top left), 0~rad~m$^{-2}$ (top right), 8~rad~m$^{-2}$ (bottom left), 15~rad~m$^{-2}$ (bottom right). The image size is $12^\circ \times 12^\circ$. The angular resolution here is 4.2~arcmin and the conversion factor is 1~mJy~beam$^{-1}$ = 1.3~K.}
\label{ncp_rm_cube}
\end{figure*}

In Figure~\ref{ncp_rm_cube} a few examples of the RM cube are shown.
%
%
Very broad arc-like structures are visible at quite different RM values with peaks in the polarized emission of 15-20~K. The largest features lie at $\alpha \sim 7^{\rm h}$, $\delta \sim 87^\circ$ and at the North-West corner of the image, but the whole field of view is crossed by similar structures at the various Faraday depths. The brightest frame of the RM cube of the 3C196 field is $\sim$2.5 times fainter than the polarized emission in the NCP area at the same Faraday depth.

From a careful visual inspection of the RM cube, we found that part of the
observed polarization might be intrinsic emission of the sky, like the emission
associated with 3C061.1. It is not possible, however, to disentagle the two
contributions at this stage and we can only conclude that the polarized emission
around the NCP is essentially due to RFI, and that the true emission of the sky is
hidden below it. Given this, we will no longer consider the polarization in the NCP area in the remainder of the paper. 

\section{Power spectrum analysis}
\label{power_spec}

In cosmological and foreground studies, diffuse emission is often analysed through its angular power spectrum, which statistically describes the spatial properties of a radiation field.

In order to characterize the power spectrum of the diffuse emission, we follow here the approach adopted in B09. We identified all the discrete sources by making sky images with only the long baselines, as we did to create a sky model. Now we did not require the CLEAN deconvolution to stop when the first negative component was found. We identified all the sources down to 30~mJy for the 3C196 field and all the sources down to 40~mJy in the NCP area.

These sources were subtracted from the visibilities and residual images were created. 

\begin{figure}
\centering
\resizebox{1.0\hsize}{!}{\includegraphics[angle=-90]{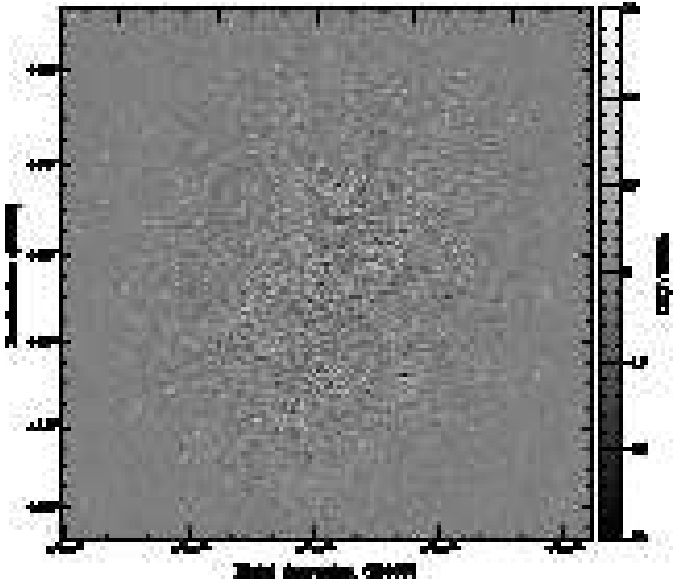}}
\caption{Stokes $I$ map of the 3C196 field after the sources down to 30~mJy have been subtracted. The conversion factor is 1~mJy~beam$^{-1}$ = 3.3~K.}
\label{3C196_res_final}
\end{figure}

\begin{figure}
\centering
\resizebox{1.0\hsize}{!}{\includegraphics[angle=-90]{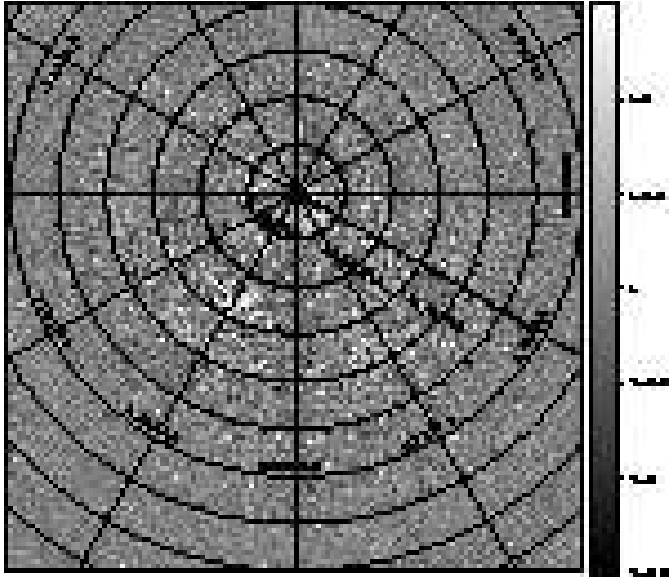}}
\caption{Stokes $I$ map of the NCP field after the sources down to 40~mJy have been subtracted. The units are Jy~beam$^{-1}$ and the conversion factor is 1~mJy~beam$^{-1}$ = 4.4~K.}
\label{ncp_res_final}
\end{figure}

Figures~\ref{3C196_res_final} and~\ref{ncp_res_final} show the $12^\circ \times 12^\circ$ residual images, where a large-scale pattern of fluctuations is now visible in both maps.
%
%
%
%
We note that no errors appear to be associated with 3C196, indicating that this source was calibrated and completely removed from the data. 

The inner $6^\circ \times 6^\circ$ quarters of the residual maps were corrected for the WSRT primary beam $A(f,\gamma)$ which was approximated as:
\begin{eqnarray}
    A(f,\gamma) = \cos^6(0.065 \, f \, \gamma)
\label{primary_beam}
\end{eqnarray}
where $f$ is the observing frequency in MHz and $\gamma$ is the angular distance
from the pointing centre in degrees. 

From the primary beam corrected maps the power spectrum was computed as (Seljak 1997, B09):
\begin{eqnarray}
   C^X_\ell = \left\{ \frac{\Omega}{N_\ell} \sum_{\bf l} X({\bf l}) X^*({\bf l}) - \frac{\Omega (\sigma^X_{\rm{noise},\ell})^2}{N_b} \right\} b^{-2}(\ell)
	\label{pow_spec_def}
\end{eqnarray}
where $X$ indicates either the total intensity $I$ or the polarized intensity $P$, $\ell=\frac{180}{\Theta}$ where $\Theta$ is the angular scale in degrees, $\Omega$ is the solid angle in radians, ${N_\ell}$ is the number of Fourier modes around a certain $\ell$ value, $X$ and $X^*$ are the Fourier transform of the image and its complex conjugate respectively, $\bf{l}$ is the two dimensional coordinate in Fourier space, $\sigma_{\rm{noise},\ell}$ is the noise per $\ell$ bin and $N_b$ is the number of independent synthesized beams in the map. The factor $b^{2}(\ell)$ is the power spectrum of the window function (Tegmark 1997). Since interferometric images represent the true sky brightness convolved with the dirty beam (the Fourier transform of the weighted $uv$ coverage), in our case the factor $b^{2}(\ell)$ is the power spectrum of the dirty beam.

The number of modes around a certain $\ell$ value depends on the bin width in Fourier space and has a minimum dictated by the width of the field of view of the instrument. The 25~m dish of the WSRT telescope gives $\ell_{min} \sim 40$, therefore we chose $\Delta \ell = 50$ as the bin width for computing the power spectrum as was done in B09.

Since the $uv$ coverage has become rather non-uniform due to the discarded baselines, the noise term $\sigma_{\rm{noise},\ell}$ may vary as a function of $\ell$. In the next section we characterize the noise from the $uv$ data rather than deriving it from the image.

\subsection{Noise estimate}
\label{noise_sim}
The rms of the noise at the angular resolution is usually determined from the map, in regions at the
edge or beyond the primary beam where no apparent emission from the sky is visible. Here we want to measure the noise by following another approach which makes use of the visibility data
alone and that might also become useful for the future actual observations of the redshifted 21~cm
line. An analogous approach has been applied by Ali, Bharadwaj \& Chengalur (2008) to the GMRT data.

We split our spectral bands in sub-bands of 100~kHz each. We assumed that in each subband the true sky,
the ionospheric effects and the calibration errors were, to a first approximation, constant. This
assumption is quite well justified for the celestial components: assuming a fiducial spectral index
for the diffuse Galactic foreground of $\beta=2.55$ (Rogers \& Bowman 2008), the signal varies by
$\sim$2\% in a 100~kHz band. For discrete sources, a spectral index  $\alpha=0.8$ causes a
variation of less than 0.1\% in a 100~kHz band. Also the ionosphere is expected not to change significantly in such a small frequency interval. 

We assumed also that the calibration does not significantly change within such a narrow band. The fact
that we achieved a very good dynamic range for the 3C196 field supports this assumption. 

\begin{figure*}
\resizebox{0.5\hsize}{!}{\includegraphics{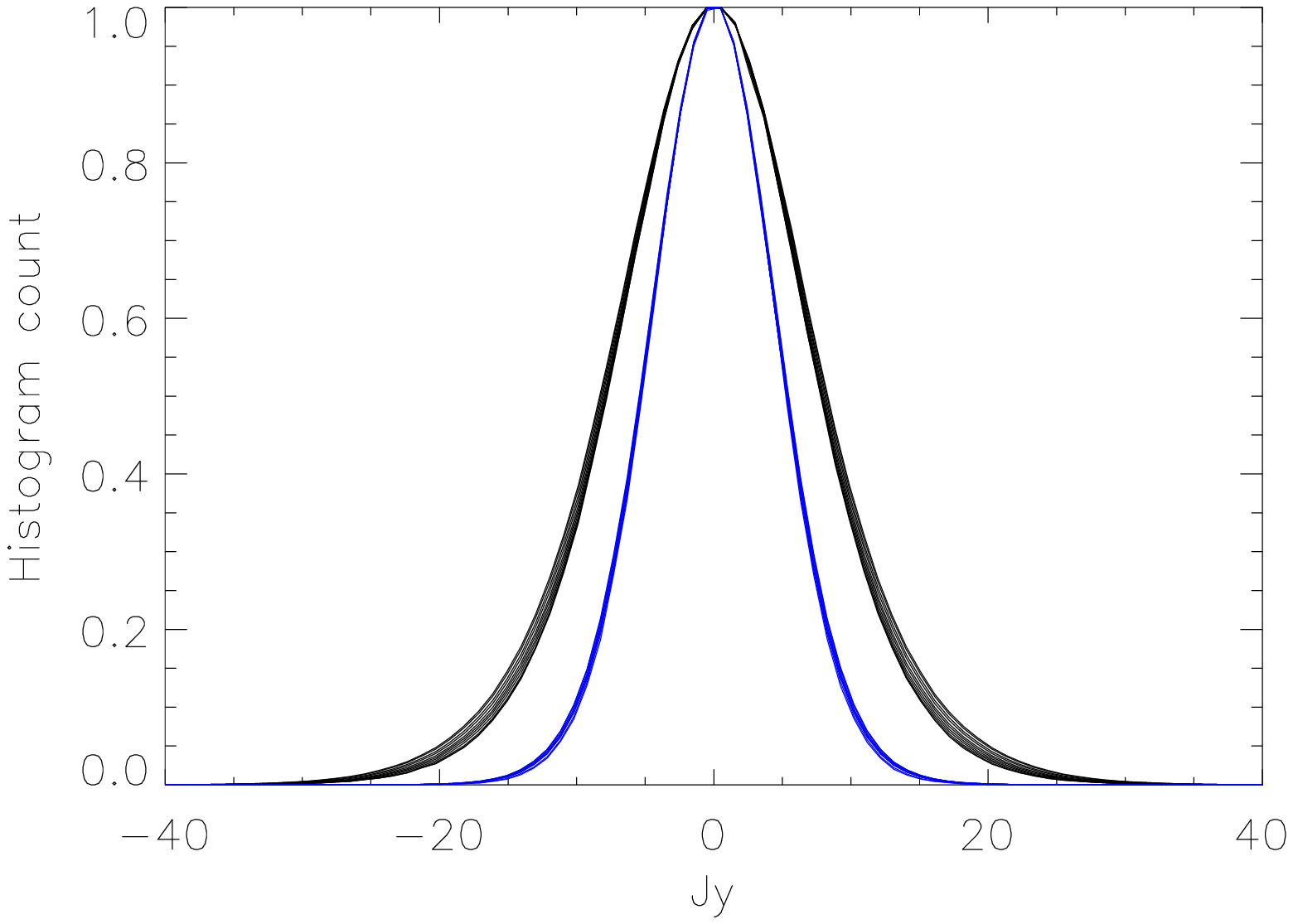}}
\resizebox{0.5\hsize}{!}{\includegraphics{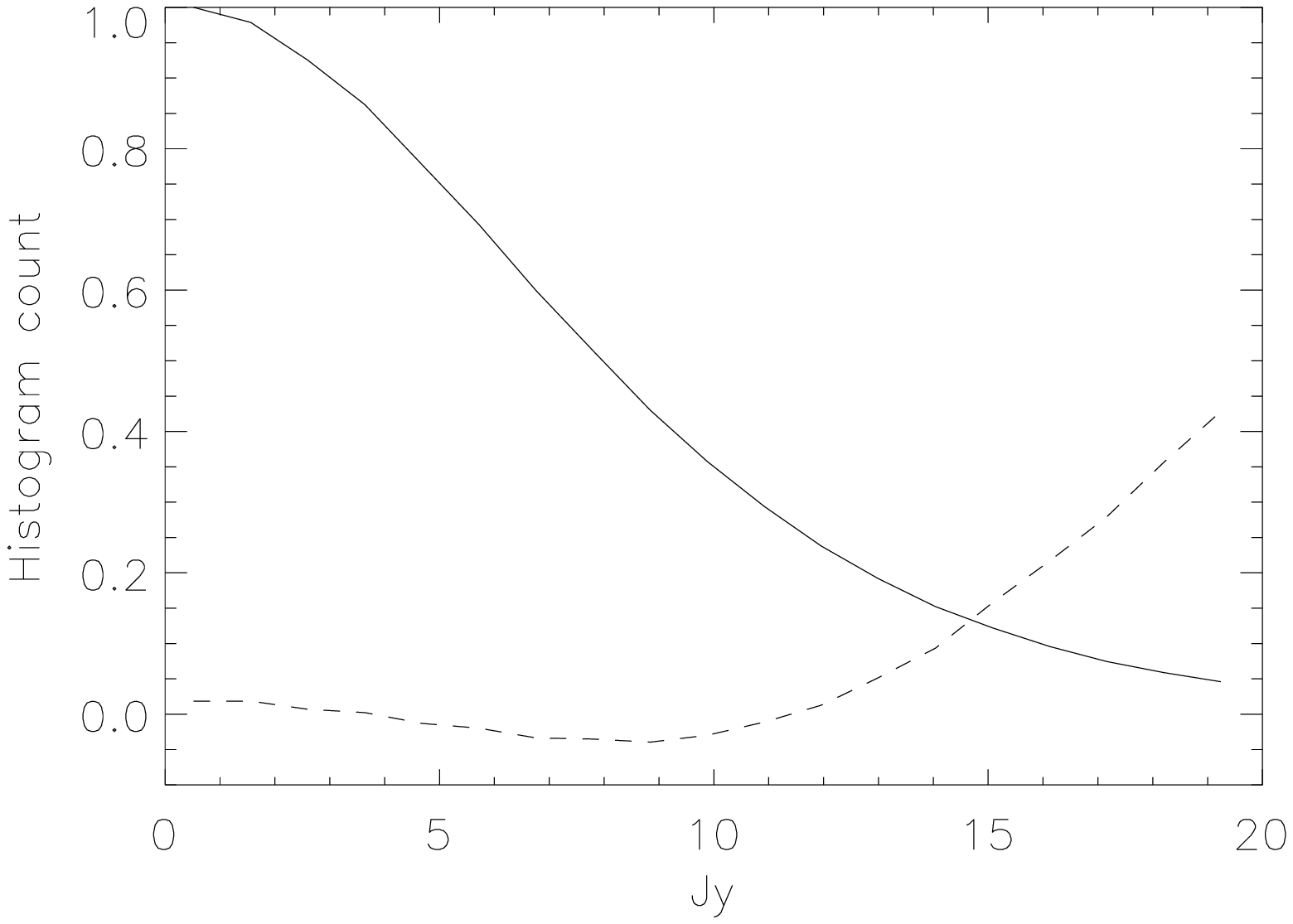}}
\caption{Left: the distribution of the real part of the residual visibilities after differencing two adjacent 100~kHz-wide subbands for the 3C196 data. The eleven samples of the 139.3~MHz (black lines) and of the 154.7~MHz (blue lines) bands are plotted. Both peaks are arbitrarly normalized to one. Right: the average distribution of the residual visibilities for the 139.3~MHz band (solid line) and the fractional residual from the best-fitting Gaussian (dashed line).}
\label{noise_histogram_3C196}
\end{figure*}

For each of the eight 2.5~MHz bands, we formed 22 sub-bands of 
100~kHz by averaging 220 of the 9.8~kHz spectral channels (in groups
of 10) from the center of the 2.5~MHz band (after discarding 26 channels at the end of the band). The 22 sub-bands were taken in pairs of adjacent sub-bands
to form 11 difference visibilities (for each polarization and time step).
Under our above assumptions, these differences should contain only instrumental
noise.  We binned the visibility data and created histograms of their
real and imaginary parts.  Fig~\ref{noise_histogram_3C196} displays examples of these distributions for the 3C196 field.

%
%

The distributions follow a Gaussian profile fairly well, as expected for instrumental random noise,
indicating that there are no major differences between two adjacent 100~kHz subbands and that this
method can provide an estimate of the noise. The fractional residual between the observed
distribution and its best-fit indicates, however, that a Gaussian profile fits the distribution reasonably well
only up to about two times the standard deviation of the distribution itself, whereas deviations from Gaussianity occur in the wings of the distributions and they can go up to $\sim$40\% at three times
the standard deviation of the distribution. A similar trend was found by Ali, Bharadwaj \& Chengalur (2008) in GMRT data.

Calibration errors and residual faint RFI can cause such deviations. In future EoR experiments,
where the cosmological signal is expected to be hidden below the noise level, it is very important to
characterize the noise properties in a very detailed way. However, for the purposes of the present
analysis, we will see that the noise measured in the image can be fairly predicted from a Gaussian distribution of the visibility noise.

It also interesting to note that the distributions related to the 139.3~MHz band are rather broader
than those related to the 154.7~MHz band, indicating that the noise is higher at 139.3~MHz. A decrease of noise as a function of frequency is expected if the sky brightness contributes significantly to the system noise. If we assume a spectral index $\beta=2.55$, we expect a $\sim$30\% change in the Galactic brightness temperature within our observing bandwidth.

\begin{figure}
\centering
\resizebox{1.0\hsize}{!}{\includegraphics{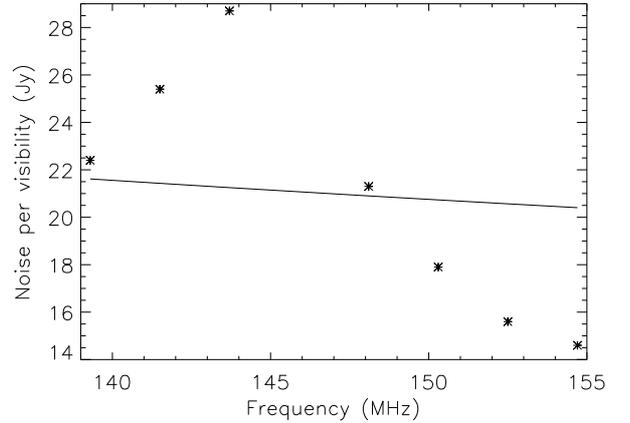}}
\caption{The measured noise per visibility as a function of frequency for the eight spectral bands of the 3C196 data (asterisks) and the behaviour of the sky dominated noise as a function of frequency (solid line). The sky dominated noise is normalized to the average noise derived from the data (see text for details).}
\label{noise_vs_freq}
\end{figure}

The noise per visibility - i.e per baseline, per time slot, per polarization and per channel - was measured as follows. The standard deviation of the best-fit Gaussian profile of every distribution gives the estimate of the noise per baseline, per time slot, per polarization per ten channels. We obtained the noise per visibility by taking the average best-fit profile of the eleven samples of noise in each band and by multiplying it by $\sqrt{10}$. We found that the noise per visibility is 22.4~Jy at 139.3~MHz and 14.6~Jy at 154.7~MHz. Converted to brightness temperature, this leads to a $\sim$62\% difference within our observing bandwidth, approximately two times bigger than expected from a pure sky dominated noise given by a $\beta=2.55$ spectrum. Figure~\ref{noise_vs_freq} displays the behaviour of the measured noise as a function of frequency.

Athough we observe a general decrease of the noise with frequency, there is a significant frequency dependent structure in the noise behaviour, indicating a more complex coupling between the expected sky brightness and the instrumental response. For instance, the noise excess in the spectral bands at 143.7 and 148.1~MHz is likely to be caused by the presence of strong RFI contamination. A detailed investigation of these effects is, however, beyond the purposes of the present work.

The average noise per visibility weighted by the number of visibilities within each band is 20.9~Jy.
We used this value to simulate a noise map which will be used later for estimating the noise power spectrum. The simulation proceeded as follows. 

For each visibility of the 3C196 data set we drew a realization of noise taken from a Gaussian distribution
with zero mean and a standard deviation of 20.9~Jy. A realization for the real and for the imaginary
part of the visibilities was drawn independently. Afterwards, a noise image was generated for each
spectral band according to the $uv$ plane weighting scheme used for imaging the sky brightness:
radial weights multiplied by a Gaussian profile with a standard deviation of 2500~m. The noise images
were then averaged to produce a final noise map. This map had an $rms$ of ~0.5~mJy~beam$^{-1}$ which
is the same value obtained from the 3C196 image in Section~\ref{calibration_3C196}.

The same procedure was repeated to estimate the noise and generate a noise map for the NCP data. We found a behaviour similar to the 3C196 data with an average noise per visibility of 31.7~Jy. The $rms$ of the simulated noise map was found to be 0.78~mJy~beam$^{-1}$ which agrees well with the value reported in Section~\ref{calibration_ncp}.
%
%

\subsection{Results}

\begin{figure}
\centering
\resizebox{1.0\hsize}{!}{\includegraphics{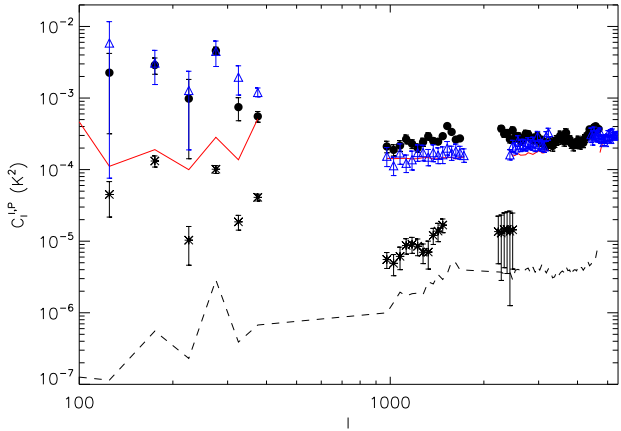}}
\caption{Power spectra of the total intensity 3C196 field (black filled
circles), of the NCP field (blue triangles), of the confusion noise (solid red
line), of the polarization in the 3C196 area (black asterisks) and of the
instrumental noise for the 3C196 data (dashed black line). The noise power
spectrum is determined through the simulations described in \ref{noise_sim}. The
error bars account for both the statistical and the instrumental noise and are
at 2$\sigma$ and 1$\sigma$ confidence level for the total intensity and the
polarization respectively. The noise has been subtracted from the total
intensity power spectra but not from the polarization power spectrum.}
\label{final_power_spectrum}
\end{figure}

The results of the power spectrum calculation are shown in Figure~\ref{final_power_spectrum} for both fields. The
error bars associated with the power spectrum were computed as (Seljak 1997):
\begin{eqnarray}
    \sigma^2_{C^X_\ell} = (C^X_\ell)^2 \frac{4 \pi}{\Omega \ell \Delta \ell} \left\{ 1+ \left[ \frac{\Omega \, \sigma^X_{\rm{noise},\ell}}{C^X_\ell b^2(\ell)} \right] \right\} ^2.
\end{eqnarray}
The two terms in the curly brackets account for the statistical and instrumental noise respectively, where the noise estimate  comes from the noise simulations presented in Section~\ref{noise_sim}.
%

Figure~\ref{final_power_spectrum} also reports the estimated power spectrum due to source confusion. If point sources have a random spatial distribution, their angular power spectrum is expected to be flat (Tegmark \& Efstathiou 1996), therefore it can be modelled as an additional noise term once its $rms$ is known. 

B09 estimated the rms confusion noise at 150~MHz to be $\sigma_{\rm{conf}} \sim 3$~mJy~beam$^{-1}$. The angular power spectrum due to the unsubtracted sources then becomes:
\begin{eqnarray}
       C_{\ell}^{\rm{conf}} = \frac{A \Omega \, \sigma^2_{\rm{conf}}}{N_b \, b^{2}(\ell)}.
\label{pow_spec_ps}
\end{eqnarray}

The factor $A$ was derived in B09 and accounts for the primary beam correction
of the WSRT. B09 showed that if a map of randomly spatially distributed
fluctuations is divided by the WSRT primary beam shape, the resulting map has an
angular power spectrum which remains flat at all angular scales, but its
normalization increased by a factor $A=1.93$. Since we corrected for the
primary beam, we accounted for this factor too.

The angular power spectra of the fields show remarkable similarities. They have both very similar intensity levels and overall behaviour. Two different regions can be seen according to their shapes.

At high $\ell$ values both spectra show a quite flat behaviour. The holes in the
power spectrum are due to the missing $uv$ coverage mentioned in
Section~\ref{obs_res}. We note that the $uv$ tracks of the NCP data are almost
circular, while the $uv$ tracks of the 3C196 field are more elongated due to the
lower declination. Therefore the hole in the power spectrum of the 3C196 field
at $\ell \sim 4000$ is partially filled by the ellipticity of the tracks.
The flat behaviour of both spectra is in good agreement with the estimated power
spectrum due to unsubtracted sources. We note that a perfect agreement is not
expected because sources were not subtracted down to the confusion level and
because the synthesized beam is slightly different for the two fields.

The exclusion of baselines 7A and 8B creates the gap at $400 < \ell < 900$ and several fluctuations in the power spectrum of the synthesized beam, as can be seen in the power spectrum of the confusion noise at $\ell < 400$. 

At low $\ell$ values both power spectra show an excess of power. Apart from the lowest multipole of the NCP power spectrum, the excess is significantly different from the power spectrum of the unsubtracted sources at 2$\sigma$ confidence level for every multipole. This excess may represent large-scale emission from the Galaxy in both fields. 

We recall that the $rms$ fluctuations $T^X_{rms}$ in a map are related to the angular power spectrum through:
\begin{eqnarray}
   T^X_{\rm{rms}}= \sqrt{\sum_{N_{\rm bin}} \Delta \ell \, \frac{\ell \, C^X_\ell \, b^2(\ell)}{2 \pi}}
\end{eqnarray}
where $N_{\rm bin}$ is the number of bins used to compute the power spectrum. After the contribution due to the confusion noise was subtracted, we found the rms of the total intensity fluctuations down to $\sim$30~arcmin to be $3.4 \pm 0.2$~K and $5.5 \pm 0.3$~K for the 3C196 and the NCP field respectively.

We tested the persistence of the excess at low $\ell$ values by computing power
spectra of the individual spectral bands and cross-power spectra of maps at two
different spectral windows. We found that the excess is present in all the bands
and correlates between different bands. 

\begin{figure}
\centering
%
%
\resizebox{1.0\hsize}{!}{\includegraphics[angle=-90]{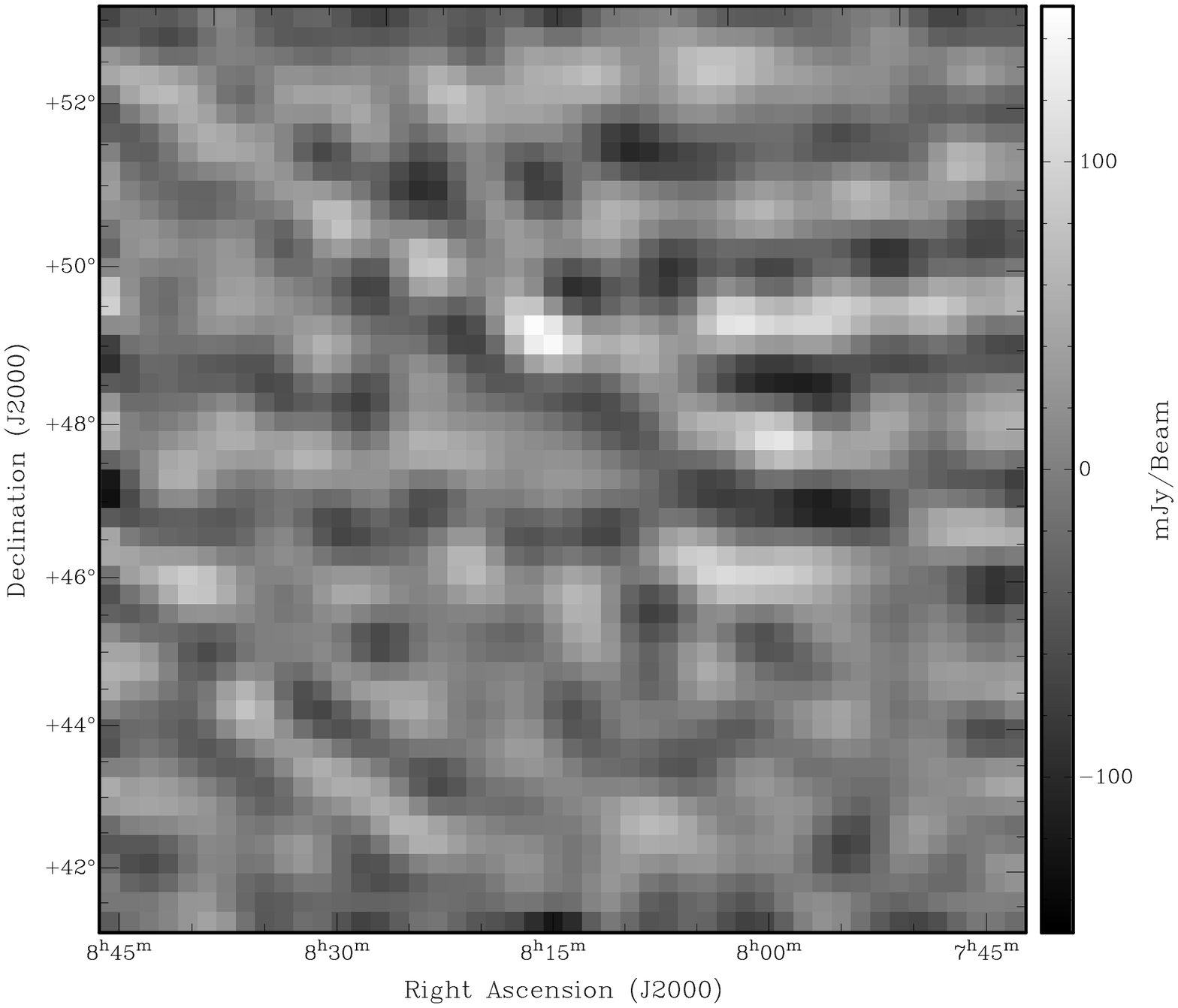}}
\resizebox{1.0\hsize}{!}{\includegraphics{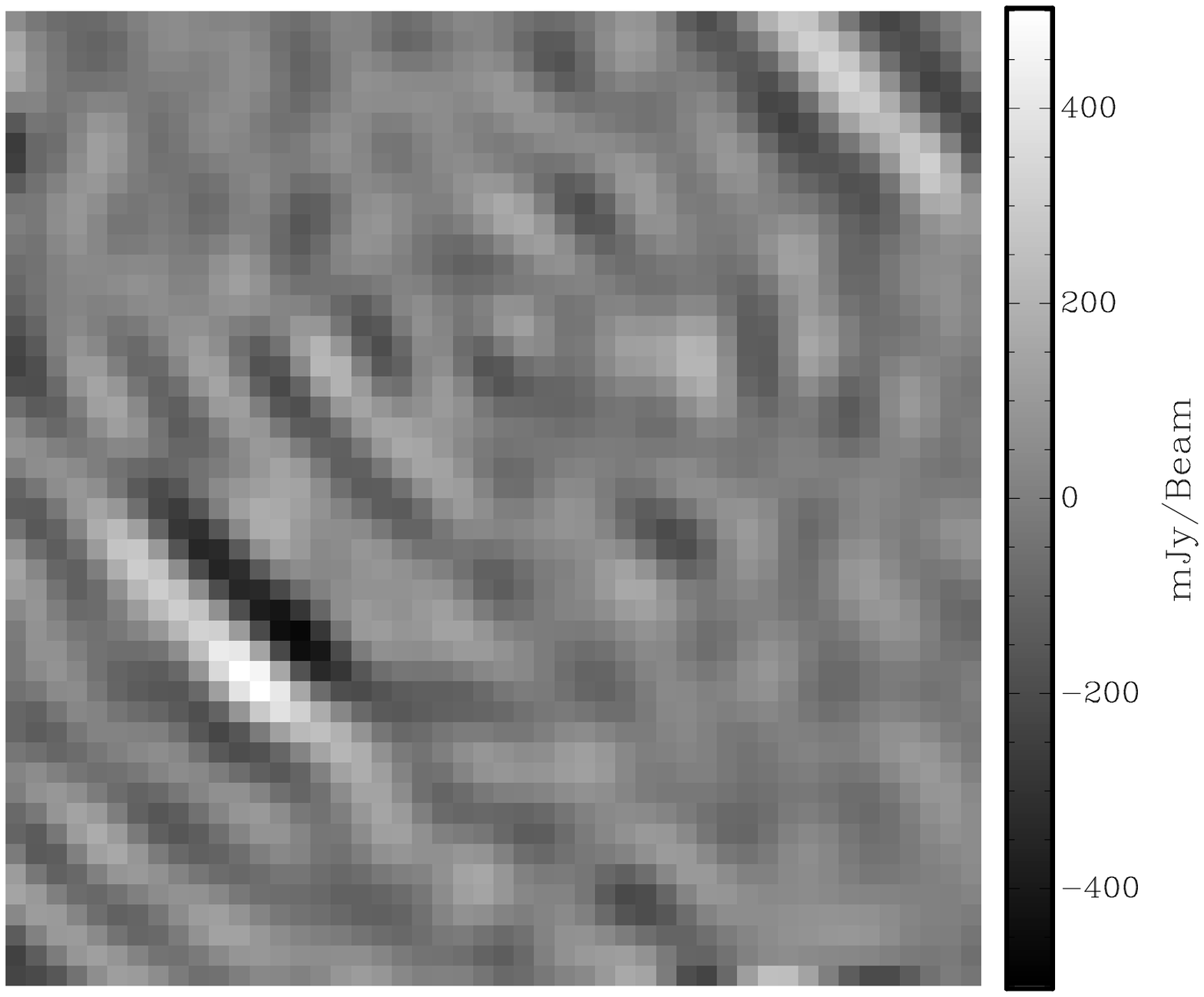}}
\caption{Top: differential imaging technique applied to the 3C196 field. The
resolution is $\sim$~44~arcmin and the conversion factor is 1~Jy
beam$^{-1}$=9.1~K. Bottom: as top but for the NCP field. The resolution is $\sim$~42~arcmin and the conversion factor is 1~Jy beam$^{-1}$=10~K.}
\label{diff_map_3C196}
\end{figure}

We performed an additional test where we looked for systematic errors in
differential images, a technique already used in the literature (Bebbington 1986). Residual sky images were created by taking a bandwith of $\sim$1~MHz and subtracting the best sky model from the visibility data. Images at adjacent frequencies were subtracted from each other so that the sky emission is cancelled out under the assumption of their smoothness in frequency. Since we are interested in the large angular scales, a taper was applied to the visibilities to lower the image resolution to $\sim$40~arcmin. The difference images were then averaged together. Figure~\ref{diff_map_3C196} shows the result for both fields.
%
%

No evident celestial sources are visible in the differential map of the 3C196
field, showing that the technique is effective in removing the sky signal. Some
side lobe residual from Cas~A is still partly visible because it was not
bright enough to be peeled and was simply modelled and subtracted from the
data. We found an rms of 0.27~K which is approximately 35\% higher than the
noise estimated from the sum of the statistical and instrumental contributions.
We conclude that systematic errors are still partly present but the excess of
power at large scales remains significant at $\sim$10$\sigma$ confidence level after including the effect of systematic errors.   

A similar conclusion holds for the NCP field. In this case the features due to systematic effects are more prominent, as could be expected from the polarization maps. The residual rms is 0.7~K, which is more than a factor two greater than the statistical noise, showing that the systematic errors actually dominate the statistical and the instrumental contributions. The excess at low multipoles is, however, still significant at $>5 \sigma$
confidence level. 

We therefore conclude that the excess at $\ell < 300$ represents large-scale emission from the Galaxy in both fields.

\begin{figure}
\centering
\resizebox{1.0\hsize}{!}{\includegraphics{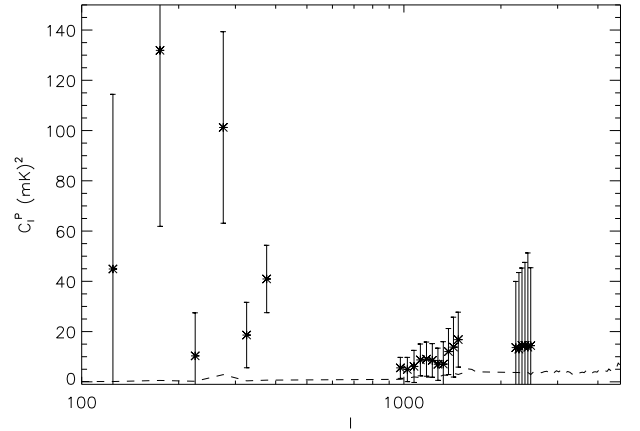}}
\caption{Polarization power spectrum in the 3C196 area (black asterisks)
and of the instrumental noise (dashed black line). The 3$\sigma$ confidence
level error bars account for both the statistical and the instrumental noise.
The noise has not been subtracted from the polarization power spectrum.}
\label{polarization_power_spectrum}
\end{figure}

Figure~\ref{polarization_power_spectrum} shows the polarization power spectrum
on a linear scale for a better comparison with the noise. It is
important to note that the noise term was not subtracted from it. 
%
%

The polarization power spectrum is more than one order of magnitude fainter than
the total intensity one and also shows a bump at $\ell < 400$. In this multipole
regime four $\ell$ values out of six are significantly different from zero at more than 3$\sigma$
confidence level. This confirms and quantifies the large-scale Galactic diffuse polarization shown in Section~\ref{pol_3C196}. 

At $\ell > 1000$ the polarization power spectrum flattens down and is consistent with the noise power
spectrum within 3$\sigma$ confidence level. We conclude that we have no significant detection of Galactic polarization on scales smaller than $\sim$10~arcmin. 
 
By integrating the power spectrum down to $\sim$30~arcmin, we found that the rms of the polarization fluctuations is $0.68 \pm 0.04$~K.

\subsection{Clustering of radio sources}

So far we have considered the confusion noise term as being due to a population of
randomly distributed radio sources. It is known that radio sources show
clustering effects (Di Matteo, Ciardi \& Miniati 2004). We have investigated whether the bump at low multipoles can be explained in terms of clustering effects of the radio sources.

In order to simulate the effect of clustering on the angular power spectrum, we
followed the approach described in Gonz\'alez-Nuevo, Toffolatti \&
Arg$\ddot{\rm{u}}$eso (2005). We outline here the steps used to create a simulated map where the point sources are clustered according to a chosen
angular correlation function. We refer the reader to their work for a full description of the method.

We assumed that sources from 15~mJy down to 10~$\mu$Jy contribute to the confusion noise observed in the data, neglecting the low flux tail of the distribution. No deep source counts are available at 150~MHz, therefore we use a $\log{N}\log{S}$ distribution derived from deep counts at 1.4~GHz (Huynh at el. 2005):
\begin{eqnarray}
   \log{\left(\frac{dN/dS}{S^{-2.5}}\right)} = \sum^6_{i=0} a_i\left[\log{\left(\frac{S}{\rm{mJy}}\right)}\right]^i
\end{eqnarray}
where $a_0=0.841$, $a_1=0.54$, $a_2=0.364$, $a_3=-0.063$, $a_4=-0.107$, $a_5=0.052$, $a_6=-0.007$, $N$ is the source count per steradian and $S$ is the flux density. According to this distribution, $\sim$140000 sources were included in our simulation. We notice that moderately deep observations down to a few mJy  at 325~MHz show similar normalized counts at 10~mJy (Owen et al. 2009), therefore uncertainties in extrapolating the source count down to low frequencies appears to be small. From the following analysis it will also be clear that the overall source count normalization is not relevant in our estimates because it is absorbed into the normalization factor that matches the simulated to the measured power spectrum.

We populated a $6^\circ \times 6^\circ$ map with 2~arcmin resolution according to a Poisson distribution of the sources. In this way we obtained a flat power spectrum for the density field of the sources. Instead of distributing the fluxes of the sources according to the $\log{N}\log{S}$ distribution as done by Gonz\'alez-Nuevo, Toffolatti \& Arg$\ddot{\rm{u}}$eso (2005), we directly normalized the angular power spectrum of the density field to the angular power spectrum of the confusion noise measured in our data. In this way we force the simulated map of point sources to match the observed flux level of the confusion noise.

The flat power spectrum of the density field can be modified according to a chosen angular correlation function $w(\theta)$, where $\theta$ is the angular scale. We assumed the angular correlation function to be (Bowman, Morales \& Hewitt, 2009):
\begin{eqnarray}
   w(\theta)=10^{-3} \theta^{-0.8},
\end{eqnarray}
where $\theta$ is measured in degrees.

The power spectrum of the correlated density field $P(k)_{corr}$ can be calculated by integrating the correlation function:
\begin{eqnarray}
   P(k)_{\rm{corr}} = \frac{2 \pi}{\Omega} \int w(\theta)J_0(k \theta) \, \theta \, d\theta
\end{eqnarray}
where $k$ is the wavenumber and $J_0$ is the zeroth-order Bessel function.

The Fourier transform of the density field can be derived from the modified power spectrum: 
\begin{eqnarray}
   \delta(\bf{k})_{\rm{corr}} = \delta(\bf{k})_{\rm{Poiss}} \frac{\sqrt{P(k)_{\rm{corr}}+P(k)_{\rm{Poiss}}}}{\sqrt{P(k)_{\rm{Poiss}}}}.
\end{eqnarray}
where $\delta(\bf{k})_{\rm{Poiss}}$ is the Fourier transform of the density field which has a Poisson distribution of the sources and $\delta(\bf{k})_{\rm{corr}}$ is the Fourier transform of the density field where the power spectrum was modified to account for the spatial correlation of the sources.
By inverse Fourier transforming we obtained the density field where the point sources are distributed according to the chosen correlation function:
\begin{eqnarray}
   \delta(\bf{x})_{\rm{corr}} = \Omega \int \delta(\bf{k})_{\rm{corr}} e^{i \, \bf{k} \cdot \bf{x}} \, d\bf{k}.
\end{eqnarray}
where $\delta(\bf{x})_{\rm{corr}}$ is the density field of correlated sources.

\begin{figure}
\centering
\resizebox{1.0\hsize}{!}{\includegraphics{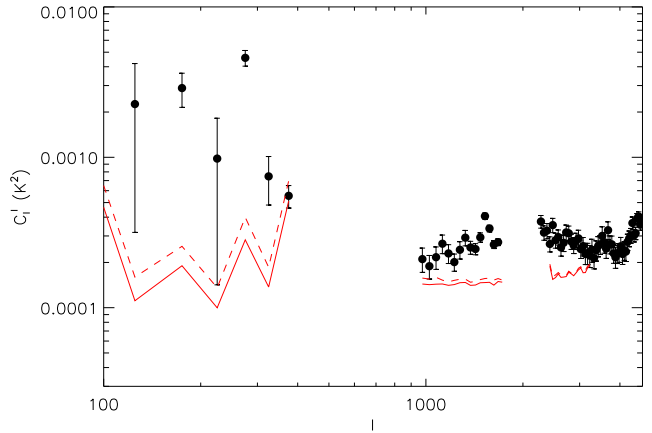}}
\caption{Comparison between angular power spectra due to various components: the 3C196 field (filled circles), Poisson distributed point sources (red solid
line) and clustered point sources (red dashed line, see text for details). The power spectra of the point sources have the same $uv$ cut as the 3C196 data. The error bars are at 2$\sigma$ confidence level.}
\label{source_clustering}
\end{figure}

We computed the angular power spectrum of the clustered density field and normalized it by the same factor used for the Poissonian density field. We generated 100 realization of the density fields in order to improve the statistics. Figure~\ref{source_clustering} shows the results of our simulations compared with the angular power spectrum of the 3C196 field.
%
%

We can see that the angular power spectra of the simulated maps show the
expected behaviour: they overlap at high $\ell$ values and the angular power
spectrum of the clustered sources increases at large angular scales. For $\ell <
300$ the difference between the clustered and the Poisson distributed point
source angular power spectra is $\sim$20\%, which is not enough to explain the
bump in the 3C196 angular power spectrum (and neither in the NCP field, where the bump is even greater). We conclude that the effect of source clustering, if present,  is not sufficient to explain the observed excess at large angular scales.

\section{Discussion and conclusions}
\label{concl}

We have presented results from observations carried out at 150~MHz with the WSRT
of two different regions at moderate Galactic latitudes, with the aim of studying
the foregrounds for EoR observations. These data represent the first investigations in sky areas where EoR observations could actually be carried out.

Data of both fields were carefully calibrated, and bright sources within the
field of view were subtracted with their own direction-dependent calibration in
order to achieve high sensitivity images. The final map of the 3C196 field
reached a thermal noise of $\sim$0.5~mJy~beam$^{-1}$ at an angular resolution of
$\sim$2~arcmin with a dynamic range of $\sim$150000:1. The final map of the NCP
field reached a thermal noise of $\sim$0.7~mJy~beam$^{-1}$ at approximately the
same angular resolution. Both images are, however, limited by confusion noise
towards the centre of the field. We found that the confusion noise is
$\sim$3~mJy~beam$^{-1}$, similar to the value found in the Fan region.

A differential method was used to estimate the instrumental noise from the
visibilities in order to derive a more accurate power spectrum of the noise. The
method gave results in good agreement with the traditional noise estimates
performed on maps and could be an important way of estimating the instrumental noise in actual EoR observations.

The angular power spectrum analysis was conducted on the residual images after the best sky models were subtracted. Both fields showed similar power spectrum behaviour in terms of both shape and magnitude. On the most relevant scales for the EoR detection -- $\theta < 10$~arcmin -- the power spectrum agrees with the contribution due to confusion noise in both fields. There is no evidence of diffuse Galactic emission on those scales.

On angular scales $\theta > 10$~arcmin our data showed systematic
errors which limited the detection of diffuse emission. After removing the corrupted baselines, the power spectrum in both fields showed an excess
at $\ell < 400$, corresponding to angular scales $\theta > 30$~arcmin. In each
power spectrum bin, this excess is 2$\sigma$ above the confusion noise even after considering the clustering effect of point sources. The rms of the Galactic signal at $\sim$30~arcmin is $3.4 \pm 0.2$~K and $5.5 \pm 0.3$~K for the 3C196 and the NCP field respectively. 

We tested the persistence of this signal in differential maps where the sky
signal was subtracted out. By subtracting adjacent images made with 1~MHz
bandwidth, the sky signal was filtered out of the data, allowing an effective investigation of residual systematic effects. We found the rms
of the differential images to be 0.27~K and 0.7~K for the 3C196 and the NCP fields
respectively. This indicates the presence of residual systematic errors, but
also excludes the possibility that they can account for the whole excess.

We concluded that this excess is due to large scale fluctuations of the diffuse Galactic emission.

The polarization data were analysed through RM synthesis. In the NCP area,
the polarization was found to be highly contaminated by RFI signals which appear
to originate at the NCP. These interferences are highly variable with time,
frequency and spatial location in the map and contaminate the RM synthesis cube
in the region $-20 < \rm{RM} < 20$~rad~m$^{-2}$ where the emission from the
Galaxy is likely to appear. They reach a brightness of a few kelvins and jeopardize the Galactic polarization. This is therefore the only upper limit
that could be placed on Galactic polarization. 

The polarization in the 3C196 field suffered from instrumental errors too. The
Stokes $U$ images could not be used due to the presence of artifacts which
look like ``whiskers'' in all the frequency channels. They are variable with time and spatial location in the maps, generating contamination in the frames of the RM cube which are interesting for detection of Galactic polarization. 
We performed an RM synthesis analysis of the polarization in the 3C196 field by using the Stokes $Q$ parameter alone. 

The RM synthesis cube showed patchy polarization in the RM frames $-3 < RM <
0$~rad~m$^{-2}$; remember that the RM cube with Stokes $Q$ alone is
symmetric around zero, therefore it has no information on the sign of the
rotation measure. This diffuse emission did not have a counterpart in total
intensity and this suggests that it originates through the Faraday screen mechanism.

The polarized power spectrum is consistent with the measured instrumental noise
for $\ell > 1000$ indicating no evidence of polarized emission on arcminute
scales. It shows a bump at angular scales greater than 30~arcmin where it is
2$\sigma$ above the noise in all the power spectrum bins. The
rms of polarization fluctuations at 30~arcmin scales is $0.68 \pm 0.04$~K. 

This result improves the upper limit of 1~K given by Pen et al. (2008) on the
polarized sky and represents the first detection of diffuse polarization at
150~MHz outside the Galactic plane. 

These results can be compared with those obtained in the Fan region in order to give a more global picture of the foreground properties at various Galactic latitudes. The Fan data cover a region just above the Galactic plane whereas the 3C196 and NCP fields are located at $b \sim 30^\circ$.

The picture emerging from these three fields is that the diffuse Galactic
foreground lacks small-scale structure and its signal is below the confusion
noise at an angular resolution of $\sim$2~arcmin. This conclusion is supported
by all three data sets. Higher resolution observations are needed to
identify and subtract the discrete sources and, therefore, reveal the Galactic
signal.

Less firm conclusions can be drawn about the Galactic foreground on intermediate and large angular scales. Galactic total intensity structure at angular scales greater than 10~arcmin was visible at low Galactic latitude but not at intermediate latitudes, where the data were affected by systematic errors.

If we limit ourselves to the largest angular scales, we observe that the emission drops by a factor 1.7-2.7 if we move from the Galactic plane to $b \sim 30^\circ$. We could not observe any power-law behaviour in the power spectrum of the Galactic signal at $b \sim 30^\circ$, but if we assume the power-law index $\beta^I_\ell=-2.2$ found in the Fan region we can extrapolate this signal to the 5~arcmin scales relevant for the EoR detection.

We find that the level of contamination at 5~arcmin in the 3C196 area could be $\delta T= \sqrt{\ell (\ell+1) C^I_\ell / 2\pi} \sim 2.2$~K. It is, however, more appropriate to take into account the full three-dimensional nature of the EoR signal in its power spectrum. 

We computed the three-dimensional power spectrum of foregrounds from the maps at each
individual frequency made after the best sky model is subtracted in order to
compare it with the theoretical expectations. We adopted the following convention:
\begin{eqnarray}
   P_k = \frac{V}{N_k} \sum_{\bf k} C({\bf k}) C^*({\bf k})
	\label{3D_pow_spec_def}
\end{eqnarray}
where $k$ is inverse of the comoving distance measured in units of {\it h}~Mpc$^{-1}$, $V$ is the volume, ${N_k}$ is the number of Fourier modes around a certain $k$ value, $C$ and $C^*$ are the 3D Fourier transform of the observed cube and its complex conjugate respectively and $\bf{k}$ is the three dimensional coordinate in Fourier space.

\begin{figure}
\centering
\resizebox{1.0\hsize}{!}{\includegraphics{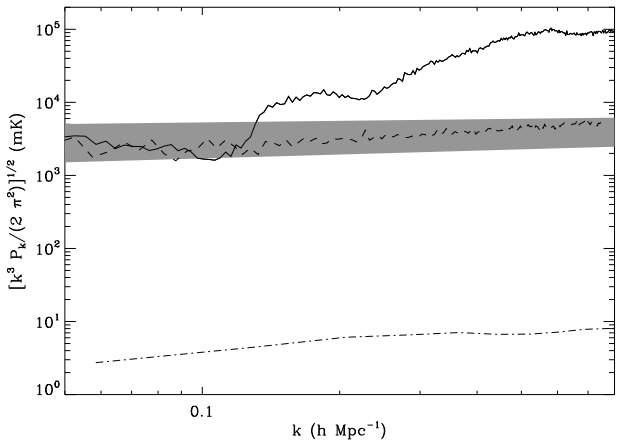}}
\caption{Comparison between power spectra due to various components: observed foregrounds in the 3C196 field (solid line), simulated diffuse foregrounds (dashed line, see text for details) and a fiducial EoR signal (dot-dashed line). The grey area accounts for uncertainties of the spatial distribution of the synchrotron emission (see text for details).}
\label{power_spectrum_3D_for_paper}
\end{figure}

Figure~\ref{power_spectrum_3D_for_paper} shows the comparison between the three
dimensional power spectrum of the 3C196 field, a fiducial EoR model and
the synchrotron diffuse foregrounds alone. We plotted the square root of the power 
spectrum which directly gives an indication of the magnitude of the various
components at that $k$ scale.

The power spectrum of the cosmological signal is obtained from the simulations
used in Harker et al. (2009a) which were extended to a box of 200~{\it h}$^{-1}$~Mpc (comoving).

The synchrotron diffuse foreground component alone was obtained by simulating a
data cube with the same observational specifications as the WSRT data. The
frequency axis contains eight bands separated by 2.5~MHz. Its statistical
properties were obtained from a Gaussian field with random phases and a
power-law distribution with $\beta^I_\ell=-2.2$ for the amplitude. The angular
power spectrum was normalized to the observed angular power spectrum of the 3C196
field for $\ell < 400$. The frequency dependence of the synchrotron component
was assumed to be $\beta=2.55$ and the spatial variation of the spectral index
was accounted for by a Gaussian distribution with standard deviation $\Delta
\beta=0.1$. If the spatial slope of the angular power spectrum changes, a change
in the slope of the three dimensional power spectrum is expected as well. The
grey area in Figure~\ref{power_spectrum_3D_for_paper} indicates the area over
which the power spectrum changes if a slope $-2 < \beta^I_\ell < -3$ is assumed.

For modes $k > 0.1$~{\it h}~Mpc$^{-1}$, point sources essentially drive the three dimensional power spectrum. At higher $k$ values the power spectrum of the unresolved sources is 1-2 orders of magnitude higher than the power spectrum due to diffuse foregrounds only. 

The power spectrum of the synchrotron diffuse foreground is approximately two
orders of magnitude greater than a fiducial EoR signal for modes $k > 0.1$~{\it h}~Mpc$^{-1}$ and is not heavily affected by changes in the spatial or spectral
slope. In regard to the observations of the cosmological 21~cm line, we
conclude that point sources are a more serious contamination than Galactic
emission on 5~arcmin scales, and that adequate observational strategies or data
analysis procedures have to be developed to face this problem.
%
%

Our data also better characterize the diffuse polarization in the Galactic halo.
We found that the polarization detected in the 3C196 field drops by a factor 
$\sim$6 if compared to the Fan area. In particular we do not see evidence of
Galactic polarization on arcmin scales where the EoR signal is expected to peak.
It is also important to note that the polarized signal at intermediate Galactic
latitude is less structured than in the proximity of the plane. The signal appears
only at a few Faraday depths, at low values of rotation measure. This is a
relevant issue for EoR observations, because small RM values generate
fluctuations on greater frequency scales and are less likely to mimic
the EoR signal in the presence of imperfect calibration. We can estimate the
magnitude of this effect considering the results from the 3C196 field.

Future low frequency arrays will have a high degree of instrumental polarization which could be 20-30\%. We assume that, with an accurate beam
model, this percentage could decrease to an average of 5\% over the whole field of view. If we also assume that the polarized signal has $|$RM$|=3$~rad~m$^{-2}$, which is
the highest value observed in the 3C196 field, the leaking signal will generate rms fluctuations of approximately 14~mK over a bandwidth of 4~MHz,
where a coherent EoR signal is expected (Morales \& Hewitt 2004). Since the cosmological signal is expected to be $\sim$5--10~mK as a function of
frequency (Mellema et al. 2006) such a contamination would pose very serious problems in extracting the EoR signal. Since the contamination is a
linear function of the polarization leakage, if this could be reduced to 1\% the contamination would become approximately a factor of two lower than
the EoR signal.

This scenario could be improved by taking a smaller bandwidth, because the Stokes $Q$ and $U$ parameters would rotate less, but this would be paid for by a decrease of the SNR.

A more effective way of mitigating this problem is to exclude the low Fourier modes where most of the Galactic
power appears, i.e. $\ell <400$. In this case, the polarization rms approaches the instrumental noise with an rms
of $\sim$0.37~K. The residual rms fluctuations would be $\sim$7.6~mK by assuming a 5\% calibration accuracy and $\sim$1.5~mK if the leakage calibration is 1\%. 

This represents only a preliminary analysis of the problem, but one which is, for the first time, supported by real data. Further
investigations of Galactic polarization in selected areas of the Galactic halo are necessary to improve the
statistics of our results, and further efforts have to be invested in the calibration of Galactic polarization and
in the removal of diffuse emission through RM synthesis. Regarding the observations of the
cosmological 21~cm line, however, Galactic polarization appears less severe than expected from the extrapolation of higher
frequency data.

\begin{acknowledgements}

We thank an anonymous referee for useful comments that helped improving the manuscript. GB thanks George Heald for useful discussions on the RM CLEAN. The Westerbork Synthesis Radio Telescope is operated by ASTRON (Netherlands Foundation for Research in Astronomy) with support from the Netherlands Foundation for Scientific Research (NWO). LOFAR is being funded by the European Union, European Regional Development Fund, and by ``Samenwerkingsverband Noord-Nederland'', EZ/KOMPAS.

\end{acknowledgements}

\end{document}